\documentclass[12pt,preprint]{aastex}
\usepackage{emulateapj5,apjfonts}
\usepackage{apjfonts}
\usepackage{graphicx}
\usepackage{amsmath}
\usepackage{color}
\usepackage{onecolfloat}

\newcommand{\lsim}{\lower0.6ex\vbox{\hbox{$ \buildrel{\textstyle <}\over{\sim}\ $}}}
\newcommand{\gsim}{\lower0.6ex\vbox{\hbox{$ \buildrel{\textstyle >}\over{\sim}\ $}}}
\newcommand{\beq}{\begin{equation}}
\newcommand{\eeq}{\end{equation}}

\newcommand{\mau}{\ h^{-1}\mathrm{M}_{\odot}}

\begin{document}
\def\head{
  \vbox to 0pt{\vss
                    \hbox to 0pt{\hskip 440pt\rm ANL-HEP-PR-11-xxxxx\hss}
                   \vskip 25pt}

\submitted{The Astrophysical Journal, submitted}

\lefthead{Bhattacharya, Habib, Heitmann, \& Vikhlinin}
\righthead{Dark Matter Halo Profiles of Massive Clusters}

\title{Dark Matter Halo Profiles of Massive Clusters: Theory vs. Observations}

\author{Suman Bhattacharya\altaffilmark{1,2}, Salman
  Habib\altaffilmark{1,2,3}, Katrin Heitmann\altaffilmark{1,2,3}, and
  Alexey Vikhlinin\altaffilmark{4,5}} 

\affil{$^1$ High Energy Physics Division, Argonne National Laboratory,
  Argonne, IL 60439, USA} 
\affil{$^2$ Kavli Institute for Cosmological Physics, The University of
  Chicago, 5640 S. Ellis Ave., Chicago, IL 60637, USA}
\affil{$^3$ Mathematics and Computer Science Division, Argonne
  National Laboratory, Argonne, IL 60439, USA} 
\affil{$^4$ Harvard-Smithsonian Center for Astrophysics, 60 Garden
  Street, Cambridge, MA 02138}
\affil{$^5$ Space Research Institute (IKI), Profsoyuznaya 84/32, Moscow,
Russia, 117997} 
 
\date{today}


\begin{abstract} 

  Dark matter-dominated cluster-scale halos act as an important
  cosmological probe and provide a key testing ground for structure
  formation theory. Focusing on their mass profiles, we have carried
  out (gravity-only) simulations of the concordance $\Lambda$CDM
  cosmology, covering a mass range of $2 \times 10^{12}-2\times
  10^{15} h^{-1}$M$_\odot$ and a redshift range of $z=0-2$, while
  satisfying the associated requirements of resolution and statistical
  control. When fitting to the Navarro-Frenk-White profile, our
  concentration-mass ($c-M$) relation differs in normalization and
  shape in comparison to previous studies that have limited statistics
  in the upper end of the mass range. We show that the flattening of
  the $c-M$ relation with redshift is naturally expressed if $c$ is
  viewed as a function of the peak height parameter, $\nu$. Unlike the
  $c-M$ relation, the slope of the $c-\nu$ relation is effectively
  constant over the redshift range $z=0-2$, while the amplitude varies
  by $\sim 30\%$ for massive clusters. This relation is, however, not
  universal: Using a simulation suite covering the allowed $w$CDM
  parameter space, we show that the $c-\nu$ relation varies by about
  $\pm$ 20\% as cosmological parameters are varied. At fixed mass, the
  $c(M)$ distribution is well-fit by a Gaussian with $\sigma_c/\langle
  c\rangle \simeq 0.33$, independent of the radius at which the
  concentration is defined, the halo dynamical state, and the
  underlying cosmology. We compare the $\Lambda$CDM predictions with
  observations of halo concentrations from strong lensing, weak
  lensing, galaxy kinematics, and X-ray data, finding good agreement
  for massive clusters ($M_{vir} > 4\times 10^{14} \mau$), but with
  some disagreements at lower masses. Because of uncertainty in
  observational systematics and modeling of baryonic physics, the
  significance of these discrepancies remains unclear.

\end{abstract}


\keywords{Cosmology: clusters-profiles  --- methods:
$N$-body simulations} } \twocolumn[\head]

\section{Introduction}
\label{section:introduction}

According to the current cosmological model, structure forms in the
Universe primarily by the amplification of primordial fluctuations
driven by the gravitational Jeans instability. The process of
nonlinear structure formation is hierarchical and complex, the initial
perturbations evolving eventually into a `cosmic web' network
consisting of voids, filaments, and clumps. The clumps, termed halos
in cosmological parlance, are dark matter dominated localized mass
overdensities with their own complex substructure. Observed baryonic
systems such as galaxies and hot gas reside in these halos. Although
the dark matter within halos cannot be observed directly, its presence
can be inferred by dynamical arguments, and much more directly,
through gravitational lensing of background sources.

The notion of the dark matter dominated halo is one of the fundamental
building blocks in studies of the formation of individual galaxies,
galaxy groups, and galaxy clusters (for an overview, see
\citealt{mo_book}). The structure of halos has been extensively
studied using N-body simulations over a wide range of halo
masses. Even though individual halos can be, and are, dynamically and
morphologically complex, it was shown by \cite{nfw1, nfw2} (NFW) that
the spherically averaged density profiles of `relaxed' halos formed in
cold dark matter (CDM) simulations can be described by a roughly
universal functional form -- the NFW profile -- independent of their
mass, the spectrum of initial fluctuations, and cosmological
parameters. The NFW profile has a fixed shape, albeit with two scale
parameters; as applied to individual halos it has been remarkably
successful and is often applied to all halos, regardless of their
dynamical state. (When applied to stacked or average halos, this
profile is somewhat less succesful, as discussed later below.)

The two parameters of the NFW profile are a mass and a scale
radius. The scale radius, $r_s$, specifies the point where the
logarithmic slope of the profile equals -2 (at small radii, the
profile $\sim 1/r$, while at large radii, it asymptotes to $\sim
1/r^3$). Instead of $r_s$, one often uses the concentration, which is
the radial scale set by the halo mass divided by $r_s$. In cluster
cosmology, the usual key observable is the halo mass, rather than the
profile per se. The cluster mass function (cluster abundance, more
generally), is a sensitive probe of dark energy, since clusters form
very late, during the epoch of dark energy dominance. However,
measuring the concentration parameter, the simplest first measurement
of a profile, can also be very useful.

First, as shown originally by NFW, the concentration of a halo, $c$,
has a strong correlation to its mass, $M$, therefore measuring the
$c-M$ relation observationally is a direct test of the CDM
paradigm. In fact, combining cluster $c-M$ predictions and
measurements, and the measured gas mass fraction, one can aim to
constrain $\Omega_m$ and $\sigma_8$ \citep{ettori11}. As another
example, lensing shear peak counts, a proposed weak lensing survey
cosmological probe, is very sensitive to the form of the $c-M$
relation \citep{king11}. Finally, future measurements of the weak
lensing power spectrum will probe small enough spatial scales that
results will be sensitive to baryonic effects on the halo profile,
i.e., modifications to the gravity-only $c-M$ relation \citep{mw04,
  zhan04}. We will return to these points in more detail below.

The correlation of halo concentration with mass is based on the idea
-- as first explicated by NFW -- that the concentration is determined
by the mean density of the universe when the halo is assembled, with
higher concentrations corresponding to higher densities. Thus cluster
mass halos, which are assembling today, should have a lower
concentration than halos of lower mass that were built up at an
earlier epoch, where the mean density was higher. Furthermore, one may
expect this trend to flatten out (sufficiently) beyond the nonlinear
mass scale $M_*$, and therefore, since $M_*$ falls rapidly with
redshift, flatten out over an extended range in mass as redshift
increases. Although the general arguments are plausible and are
broadly consistent with simulation results, a predictive theory for
the mean of the $c-M$ relation, and its scatter, does not
exist. Several simple heuristic models tuned to simulations have been
suggested (NFW; \citealt{bullock99}; \citealt{eke01};
\citealt{zhao09}) but their predictive status cannot be considered
satisfactory, especially at the higher end of halo masses (see, e.g.,
\citealt{gao07}; \citealt{hayashi07}; \citealt{maccio08};
\citealt{zhao09}). Indeed there is sufficient uncertainty even when
comparing simulation results from different groups, that the general
problem is still open. However, as the mass resolution in large-volume
simulations continues to improve, we may expect this situation to be
merely temporary.

On the observational front, cluster (and group) halo profiles can be
studied using both strong and weak gravitational lensing,
individually, and in combination (see, e.g., \citealt{comerford07};
\citealt{broadhurst08}; \citealt{mandelbaum08}; \citealt{okabe09};
\citealt{oguri11}; \citealt{zitrin11}; \citealt{coe12}), projected gas
density and temperature profiles from X-ray observations (see, e.g.,
\citealt{vikhlinin05}; \citealt{buote06}; \citealt{schmidt06};
\citealt{gastaldello07}; \citealt{vikhlinin09}; \citealt{sun09};
\citealt{ettori11}), and galaxy kinematics (\citealt{diaferio05};
\citealt{rines06}; \citealt{wojtak10} and references therein). Results
from these observations have generally shown qualitative agreement
with the $c-M$ relation obtained from simulations, although there have
been difficulties with matching the shape and
normalization. Additionally, there are discrepancies between different
sets of observations, presumably because the underlying systematics
are not fully understood and modeled.

The purpose of this paper is to present a set of predictions for the
NFW mass profile targeted primarily towards massive clusters. To do
so, however, a fairly large mass range must be considered in order to
obtain a sufficiently well-defined $c-M$ relation. Our simulations
cover three orders of magnitude in mass ($\sim 10^{12}-\sim 10^{15}
h^{-1}$M$_\odot$) with very good control of statistics over the entire
range. The high dynamic range and excellent statistics enable us to
derive a new set of results for the mass profile, including profile
evolution and probability distribution functions (PDFs) for the
concentration as a function of mass. We compare our results with
previous simulations and with a set of recent observations of the
cluster mass profile.

The paper is organized as follows. In Section~\ref{section:halo} we
discuss general features of the $c-M$ relation in the simulation
context focusing on the role of differing definitions and analyses. In
Section~\ref{section:sims}, we describe the main features of the
simulation runs. We present our results for the $c-M$ relation and its
redshift evolution in Section~\ref{section:results}. This is followed
(Section~\ref{sec:wcdm}) by a presentation of results from a suite of
$w$CDM cosmologies in order to further study how the concentration
depends on cosmology. Next, in Section~\ref{sec:comp}, we provide
a detailed comparison with recent observations, noting areas of
agreement and disagreement. Finally, Section~\ref{section:disc} is
devoted to a summary of the results and further discussion. An
Appendix discusses various systematic issues that need to be
considered when deriving concentrations from simulation results. A
number of tests are used to illustrate these points and to verify the
robustness of the numerical procedures carried out in this paper.

\section{Halos and Concentrations}
\label{section:halo}

Dark matter dominated halos are dynamically complicated and rendering
them as simplified `few parameter' objects involves a fair degree of
approximation, opening the possibility of biases in the sense that
different procedures will inevitably yield different results -- what
these different results may imply for observations is yet another
question. In this paper we adopt a minimal approach to describing
halos; we consider the first approximate description of a halo to be a
spherically averaged profile with a single power law and one overall
parameter (e.g., singular isothermal sphere), and the NFW profile as
essentially taking the next step with a broken power law and two
parameters (the mass and the concentration). In three dimensions,
halos are known to be triaxial with a major axis roughly twice as long
as the two minor axes (roughly equal)~\citep{jing02}. Spherical
averaging of this profile yields the NFW broken power-law.

In reality, halos have complicated substructure and complex infall
regions, all of which may make interpreting the concentration somewhat
nontrivial, as well as introduce projection-related biases in
observations (e.g., \citealt{white10}). Nevertheless, as shown by
\cite{evrard08}, cluster-scale systems with mass greater than
$10^{14}h^{-1}$M$_{\odot}$ are dominated by large, primary halos --
satellite halos carrying only $\sim 10\%$ of the mass -- and possess a
well-defined and regular virial relation. Therefore, it appears
reasonable to proceed in the manner outlined above.

The lack of smoothness in the individual radial density profiles of
halos -- even at high mass resolution -- means that the simple NFW
description will have varying levels of success (see, e.g.,
\citealt{tormen96}; \citealt{lukic09}; \citealt{reed11}) on a halo by
halo basis. Average or stacked profiles are of course much smoother;
it turns out that such profiles systematically deviate from the NFW
form and another scale parameter is often introduced to improve the
fit, leading to the so-called Einasto profile (see, e.g.,
\citealt{gao07}). While this improves the stacked fit primarily at
smaller radii, it has little effect on measurements of the
concentration for individual halos \citep{gao07, reed11}. Since our
objective is to carry out comparisons primarily against observations
of individual objects, rather than against correlation functions or
stacked observations, we do not use the Einasto profile.

An important piece of missing physics in our simulations is the lack
of non-gravitational baryonic effects. This is a very difficult
problem to deal with for galaxy and group-scale objects, but less so
for clusters. In clusters, the dominant form of atomic matter is not
stars, but hot gas. Gas cooling does not have a major effect on the
profiles except close to the inner regions of the cluster, roughly
$r<0.1R_{vir}$ \citep{kazantzidis04, duffy10, cui11}. Beyond this
radius the gas distribution is determined by the self-consistent
gravitational potential. \cite{duffy10} have carried out an extensive
study of possible baryonic effects (cooling, feedback) on cluster
profiles and concluded that the baryonic effects are likely to alter
the concentration at most at the $10\%$ level. This is roughly the
level of systematic control over the current gravity-only measurements
of halo concentrations, therefore we do not concern ourselves with
estimating baryonic effects or trying to correct for them, beyond not
fitting for the concentration at radii, $r<0.1R_{vir}$. The fact that
we have good agreement with observations for massive clusters
(Section~\ref{sec:comp}) may be viewed as added support to the
argument that baryonic effects do not influence cluster profiles away
from the inner regions.

A large number of numerical studies have been carried out
investigating halo profiles and paying close attention to the behavior
of the density cusp on the very smallest scales. We are, however,
concerned not with these scales, but more with scales of order
$\sim(0.1-1)R_{vir}$, since our target halos have relatively modest
concentrations. Previous numerical simulations have found that in the
region of interest to us, the concentration is a slowly varying
function of mass, typically described by power laws with index
$\alpha\simeq-0.1$ at $z=0$. These simulations have varied widely in
dynamic range, box size, and mass resolution. Partly as a result of
this, there have been some disagreements in the value of the slope and
the normalization of the $c-M$ relation, and also some lack of clarity
regarding the reasons underlying the differences.

Among the more recent studies are those involving the Millenium
simulation (MS) \citep{springel_ms} with $2160^3$ particles and a box
of side 500 $h^{-1}$Mpc assuming a WMAP1 cosmology \citep{neto07,
  gao07, hayashi07}. Halo profiles were investigated over a mass range
of $10^{12}-10^{15} h^{-1}$M$_\odot$ and it was found that
$\alpha\simeq-0.1$. These results were mostly in agreement with a
simulation campaign conducted by \cite{maccio08} and \cite{maccio07}
who covered a mass range $10^9-10^{13}h^{-1}$M$_\odot$, although with
a slight discrepancy ($\sim$ 10\%) in the
normalization. \cite{duffy08} carried out another set of simulations
with three different box sizes (25, 100 and 400 $h^{-1}$ Mpc), each
with $512^3$ particles covering a mass range of $10^{11}-10^{15}
h^{-1}$M$_\odot$ using the best-fit WMAP5 cosmology. They concluded
that the median $c-M$ relation is lower by about 23\% at the low mass
end and 16\% at the high mass end compared to the MS results in the
mass range of $10^{11}-10^{14} h^{-1}$M$_\odot$. In yet another set of
simulations, \cite{klypin10} and \cite{prada11}, have claimed that the
concentration, instead of flattening out at high mass, in fact rises.

Given this context, our primary purpose is to improve the statistical
power in determining the $c-M$ relation and its scatter at high
masses, while retaining good mass resolution, and second, to study the
behavior as a function of redshift and cosmology. Finally, we note
that the improved statistical power is important in comparing with
observations of massive clusters as the numbers of well-observed
clusters is expected to rise significantly in the near future (in the
past, simulations may have contained only one cluster at the upper
mass end, where we have hundreds).

\section{Simulation Suite}
\label{section:sims}

Throughout this paper, we use the following $\Lambda$CDM cosmology as
a reference: $\omega_m=0.1296$ ($\Omega_m=0.25$), $\omega_b=0.0224$
($\Omega_b=0.043$), $n_s=0.97$, $\sigma_8=0.8$, and $h=0.72$ where
$\omega=\Omega h^2$ and $\Omega_m$ represents the total (dark +
baryon) matter density. We assume spatial flatness. This model is in
excellent agreement with the latest best-fit cosmological model
provided by WMAP-7 measurements~\citep{wmap7}.  In order to cover a
wide range of masses, we analyze three simulations with different
volumes and number of particles. A summary of the runs is given in
Table~\ref{tab1}. The mass resolution in the large-box run is
sufficient for measuring the concentrations for halo masses
$>10^{14}h^{-1}$M$_\odot$, with a minimum of 2000 particles per
halo. At $z=0$, we have more than 100,000 such halos, therefore our
statistical control may be considered to be more than satisfactory. In
the MS and \cite{duffy08} simulations, the largest boxes used are of
size $500h^{-1}$Mpc and $400 h^{-1}$Mpc respectively, with limited
statistics for cluster size halos in the mass range $10^{14}-10^{15}
h^{-1}$M$_{\odot}$. We provide a large sample of cluster size halos,
with roughly 64 times more volume than in the MS run and 125 times
more than in the simulations by \cite{duffy08}.

\begin{table}
\begin{center} 
\caption{Description of the simulation suite}
\label{tab1}
\begin{tabular}{ccccc}
Code &  Box & Softening & Particles & $m_p$\\
\hfill  &  [$h^{-1}$Mpc] & [$h^{-1}$kpc] & \hfill &  [$h^{-1}$M$_\odot$]\\
\hline\hline
{HACC} (HACC) &2000 & 7 &2048$^3$ & $6.5\cdot 10^{10}$\\
{HACC} (HS) &512 & 7 &2048$^3$ & $1.1\cdot 10^{9}$\\
{\sc GADGET-2} (G)& 936 & 36 &1024$^3$ & $5.3\cdot 10^{10}$\\
{\sc GADGET-2} (GS)& 128 & 10 & 512$^3$ & $1.1\cdot 10^{9}$\\
\hline
\vspace{-1.5cm}
\tablecomments{Runs are referred to in the paper by
  the names in parantheses. HS \newline was run up to $z=2$.}
\end{tabular}
\end{center}
\end{table}

The largest simulation (both with respect to volume and particle
number) is carried out using our new Hardware/Hybrid Accelerated
Cosmology Code (HACC) framework described in \cite{habib09} and
\cite{pope10}. This simulation covers a volume of (2~$h^{-1}$Gpc)$^3$
and evolves 2048$^3$ particles and was run on the hybrid supercomputer
Cerrillos at Los Alamos National Laboratory. (Another 2048$^3$
particle run with a 512 $h^{-1}$Mpc box was used to test the results
obtained at $z=2$ from the {\sc GADGET-2} run described below.) The
HACC framework has been designed with flexibility as a prime
requirement; it is meant to be easily portable between
high-performance computing platforms based on different
architectures. The first version of the code has been optimized to run
on the Cell-hybrid architecture shared by Roadrunner (the first
computer to break the Petaflop barrier) and Cerrillos. A first
extension of this version of the code has been developed for hybrid
CPU/GPU systems, written in OpenCL.

HACC's code structure is split into two components: a long-range force
solver and a short range module. The long-range force solver uses a
parallel Particle-Mesh (PM) algorithm with spectral filtering and
super-Lanczos differentiation \citep{hamming}. In this part of the
code, the long-range force is calculated by depositing tracer
particles onto a regular grid and using Fourier transform methods to
solve the Poisson equation (with in effect a modified Green function)
and then interpolating the force from the grid back onto the
particles. The spectral component of the code is implemented in
C++/MPI and can run on any standard parallel machine. The current 2-D
domain-decomposed implementation of the FFT allows it scale to
millions of MPI ranks.  The implementation of the particle deposition
and force interpolation routines depends on the machine
architecture. On Roadrunner and Cerrillos, these routines were
implemented on the Cell processor.

The short-range module adds the high-resolution force between
particles and can be implemented in different ways and on different
platforms. On Cell and GPU-based systems, an $N^2$-algorithm is used
to evaluate the short range forces (in chaining mesh patches), leading
to a P$^3$M implementation. This works well on hardware-accelerated
machines since it is computationally intensive and uses a simple data
structure. The P$^3$M version of HACC has been extensively tested
against the code comparison suite results of \cite{heitmann05}. On
non-heterogeneous systems, such as the IBM BG/Q, the $N^2$ algorithm
is replaced by a recursive coordinate bisection (RCB) tree method to
guarantee good performance.

In addition to the main code base, a parallel analysis framework for
HACC has been developed. This framework runs on conventional
supercomputing hardware (or on the `top' layer of a heterogeneous
system). Among other utilities, it contains a halo and sub-halo
finder. The halo finder was part of a recent comparison project
\citep{knebe11} and is used for the analysis results presented in this
paper. Major parts of the HACC analysis framework have been
implemented into ParaView and publicly released \citep{woodring11}.

The smaller simulations are carried out with {\sc GADGET-2}, a
publicly available TreePM code~\citep{springel05}. Of these, the
larger simulation -- (936~$h^{-1}$Mpc)$^3$ volume, 1024$^3$ particles
-- is part of the Coyote Universe simulation suite \citep{heitmann10,
  heitmann09, lawrence10} which spans 38 $w$CDM cosmologies. This
simulation was also used to derive a high-precision $\Lambda$CDM mass
function prediction~\citep{bhattacharya11}. (The HACC mass function in
the large-volume run presented here is in excellent agreement with
these results). The smallest of the three simulations --
(128~$h^{-1}$Mpc)$^3$ volume, 512$^3$ particles -- serves three
purposes: (i) It allows us to probe halos at small masses, (ii) it
provides large overlap with previous work and therefore connects our
new results to a mass range that has been extensively studied in the
past, and (iii) because it is run with a completely different code, it
provides an excellent check on possible code-related systematics (for
which we find no evidence -- more details are in the Appendix).

The initial conditions for all three simulations are generated using
the Code for Anisotropies in the Microwave Background (CAMB
\footnote{http://camb.info}) and the Zel'dovich approximation at a
high starting redshift, $z_i\simeq 200$. Further discussions regarding
simulation accuracy issues can be found in the Appendix.

\section{$\Lambda$CDM Results}
\label{section:results}

\subsection{$c-M$ relation}

\begin{figure*}
  \begin{center}
    \begin{tabular}{cc}
      \resizebox{3.6in}{3.6in}{\includegraphics{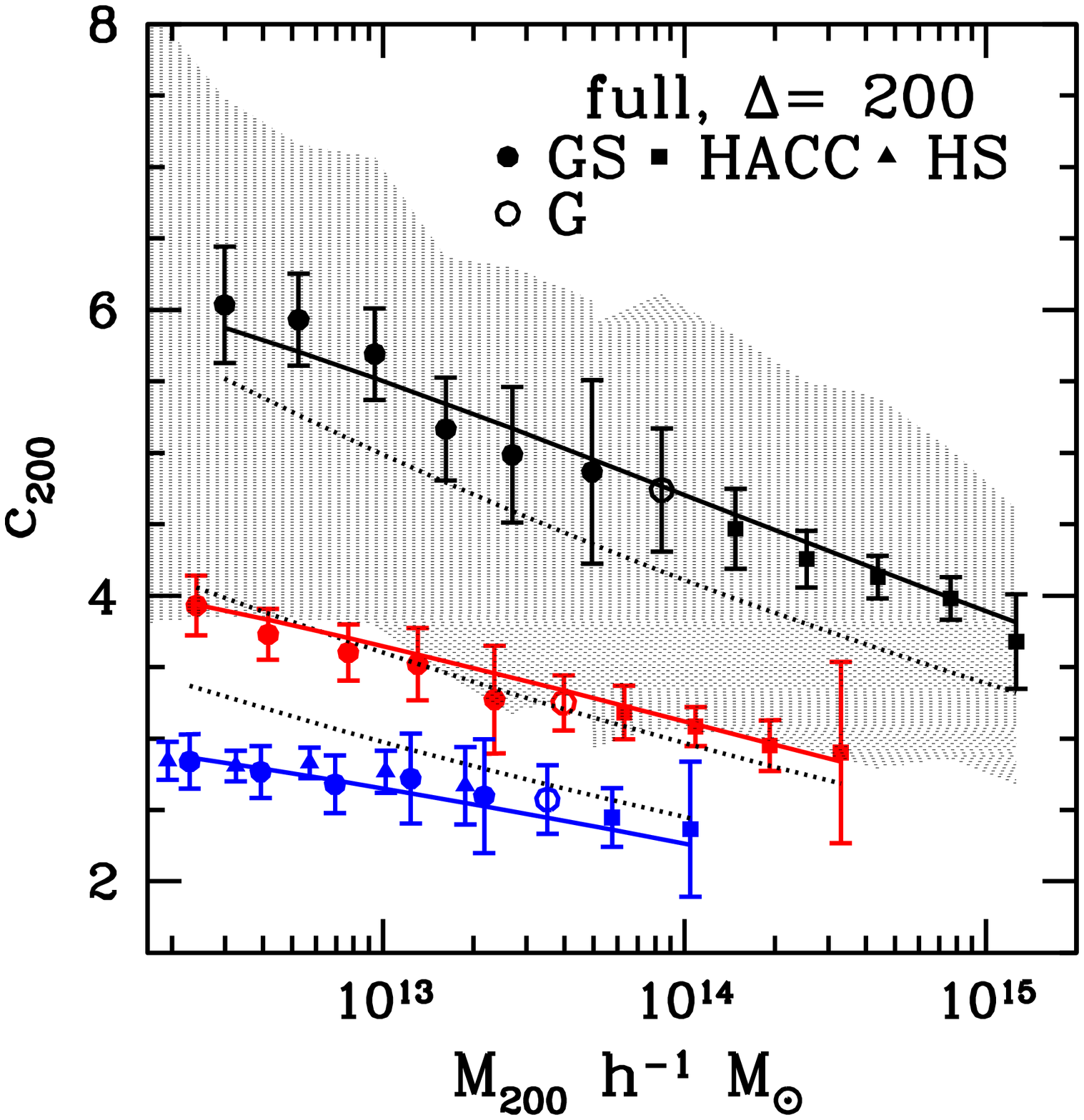}}
         \resizebox{3.6in}{3.6in}{\includegraphics{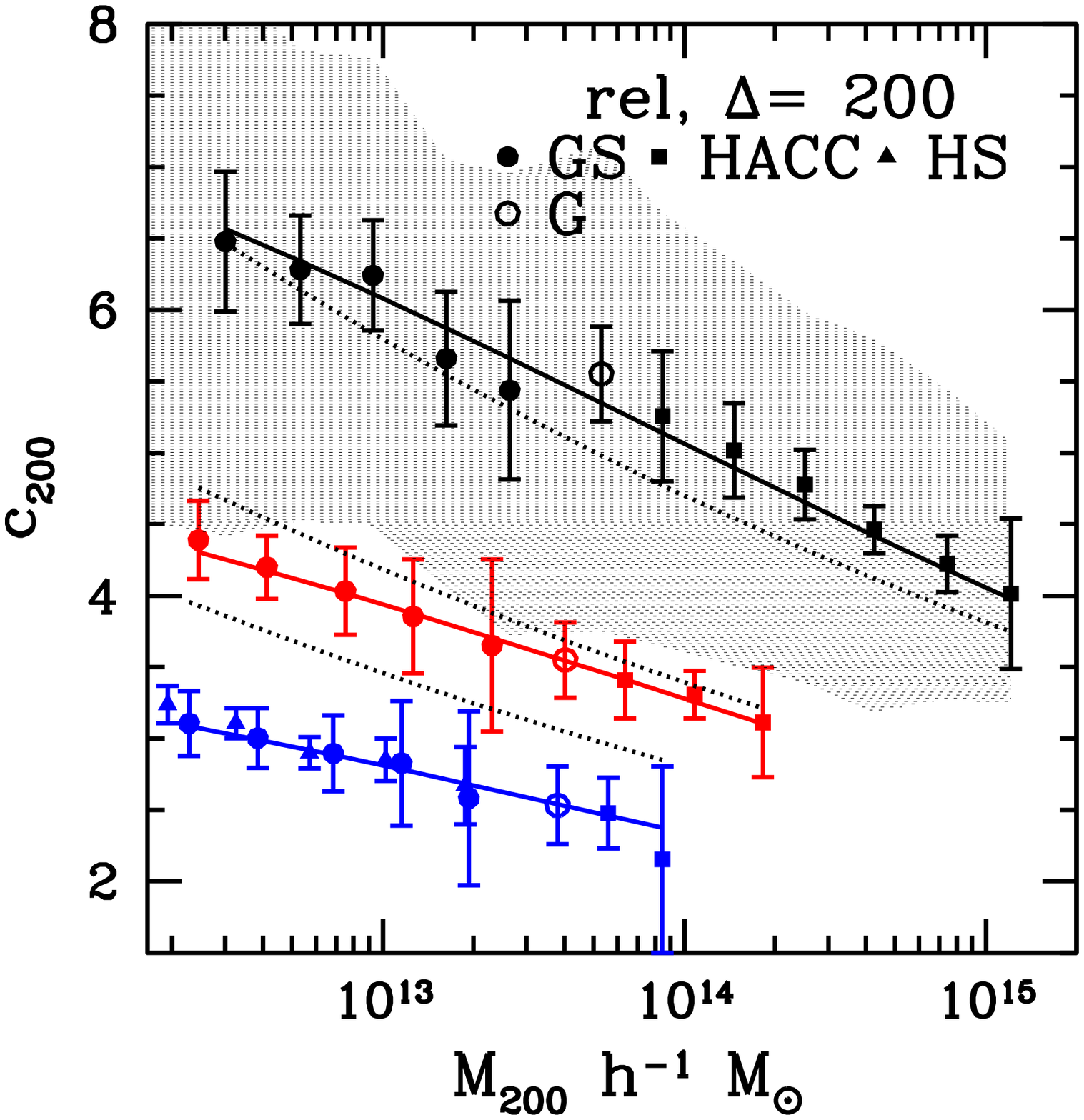}}\\
        \resizebox{3.6in}{3.6in}{\includegraphics{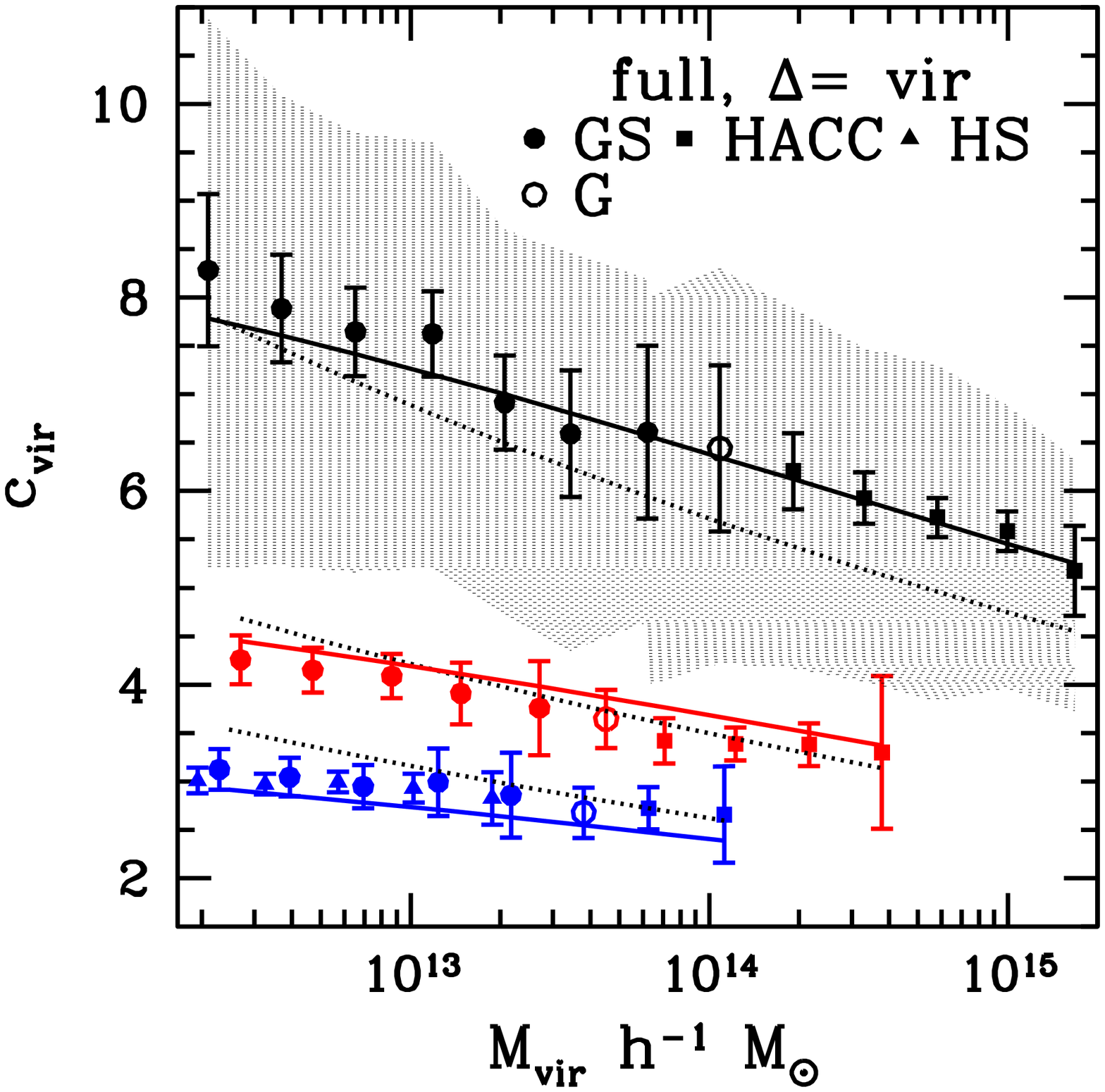}} 
        \resizebox{3.6in}{3.6in}{\includegraphics{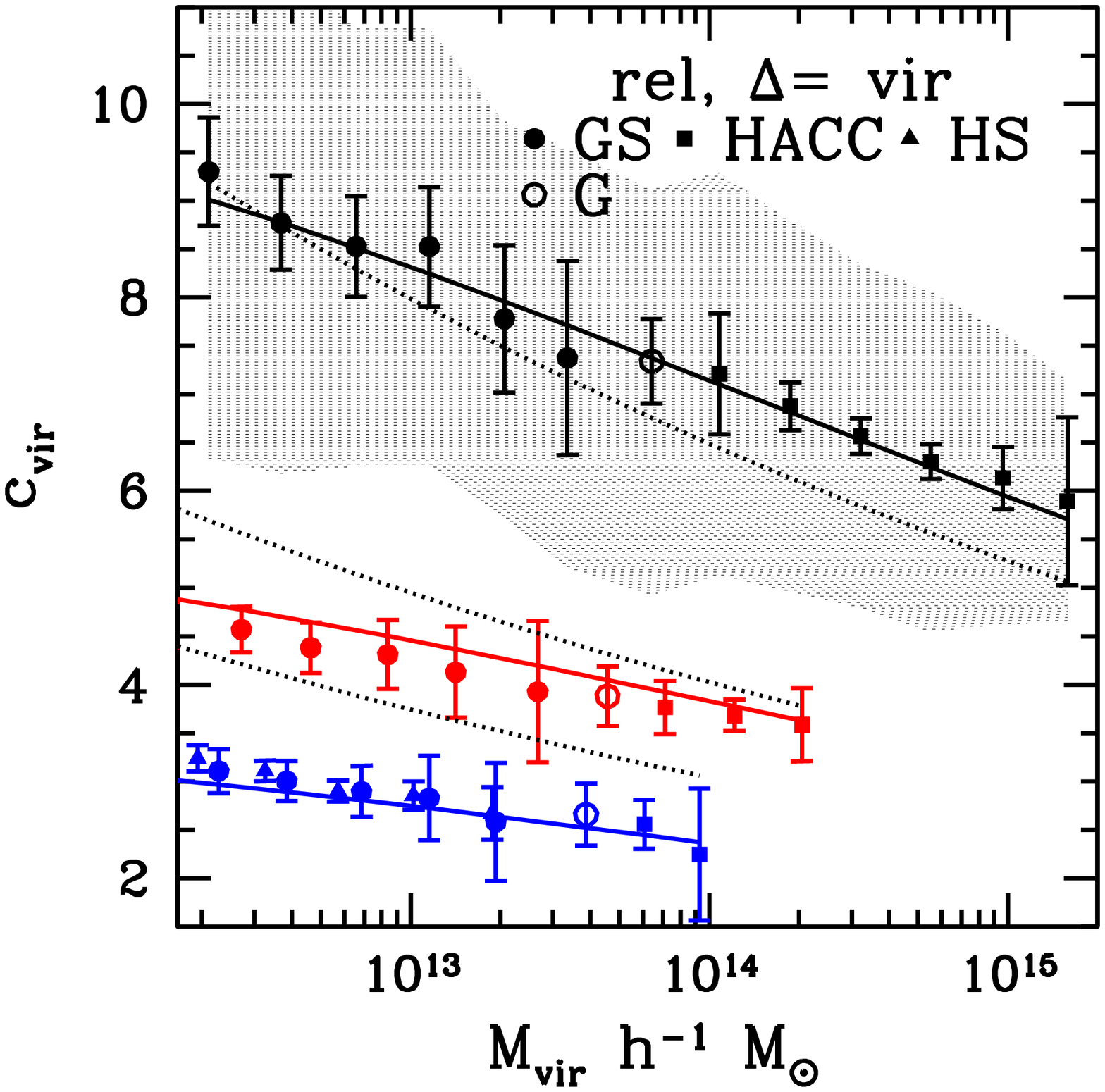}} 
        \end{tabular}
        \caption{$c-M$ relations at radii $r=R_{200}$ and $r=R_{vir}$
          for $z=0$ (black), 1 (red), and 2 (blue) for the full (left
          panels) and relaxed samples (right panels), combining
          results of multiple simulations. The black solid lines at
          $z=0$ are power law fits, $\alpha=-0.08$ for the full sample
          and $\alpha=-0.084$ for the relaxed halos. The solid red and
          blue curves are from the global fit (across all redshifts)
          discussed in Section~\ref{subsec:c_nu} and shown in
          Fig.~\ref{fig:c-Mnu}. The error bars represent the error in
          determining the mean of the concentration in each mass
          bin. At a given mass, the distribution of concentrations is
          Gaussian with standard deviation $\sigma_c/c\sim1/3$
          (Cf. Section~\ref{sub:concdist}) -- the shaded region shows
          the $1\sigma$ boundary for $z=0$. The dotted curves are
          fitting formulae for the median concentration as given by
          \cite{duffy08}.}
\label{fig:c-M}
  \end{center}
\end{figure*}

\begin{figure*}
  \begin{center}
    \begin{tabular}{cc}
      \resizebox{3.6in}{3.6in}{\includegraphics{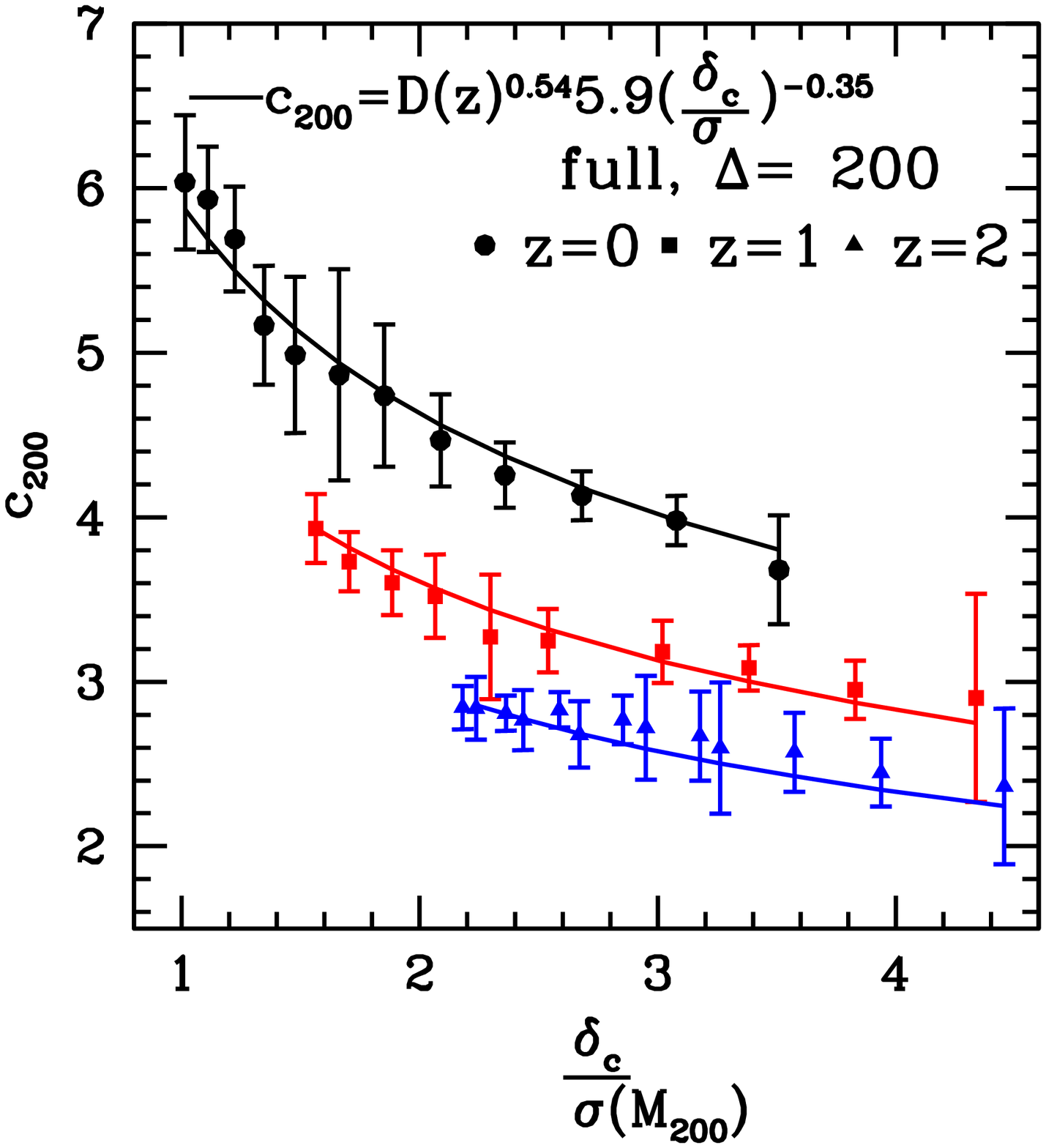}}
         \resizebox{3.6in}{3.6in}{\includegraphics{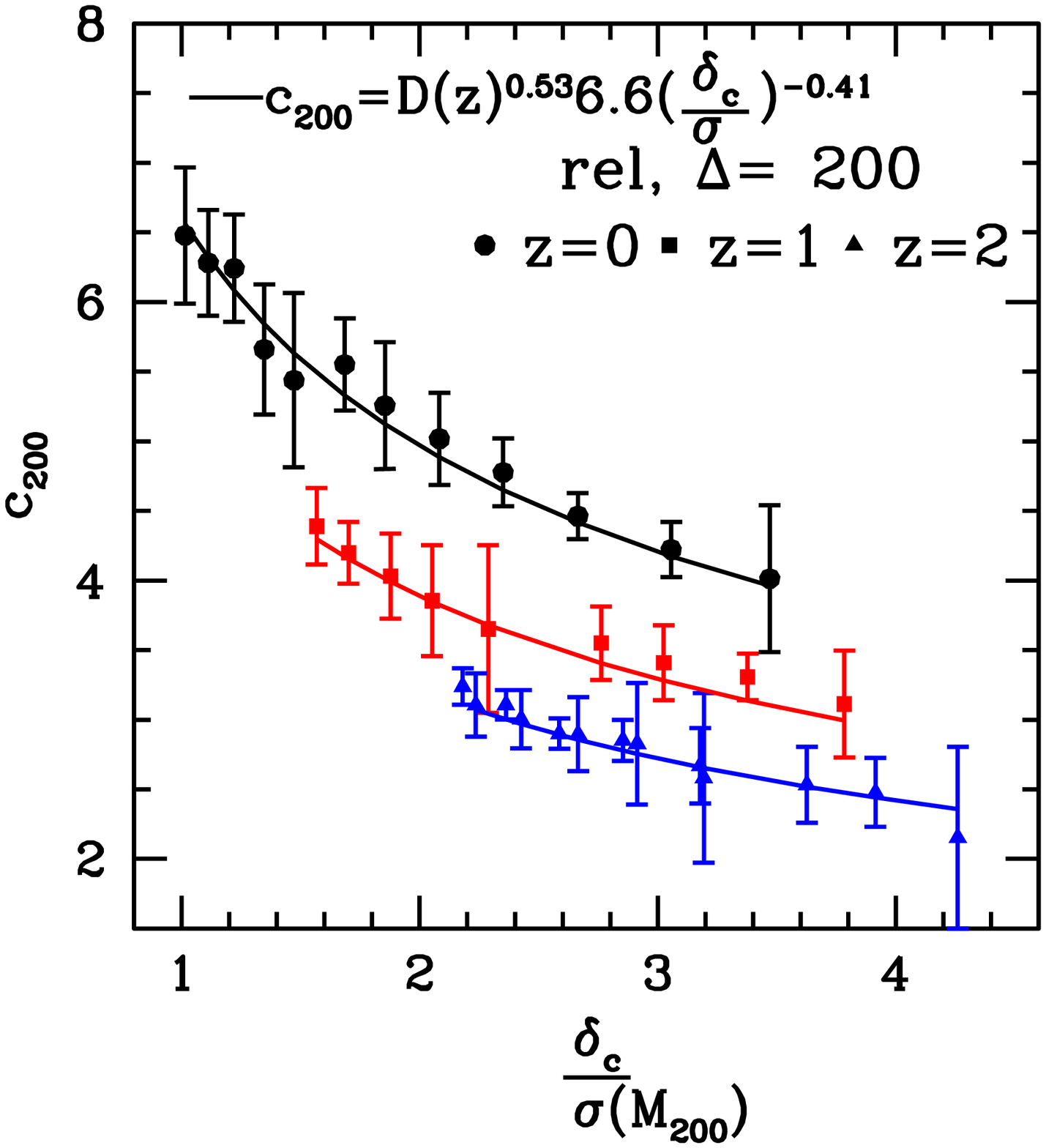}}\\
        \resizebox{3.6in}{3.6in}{\includegraphics{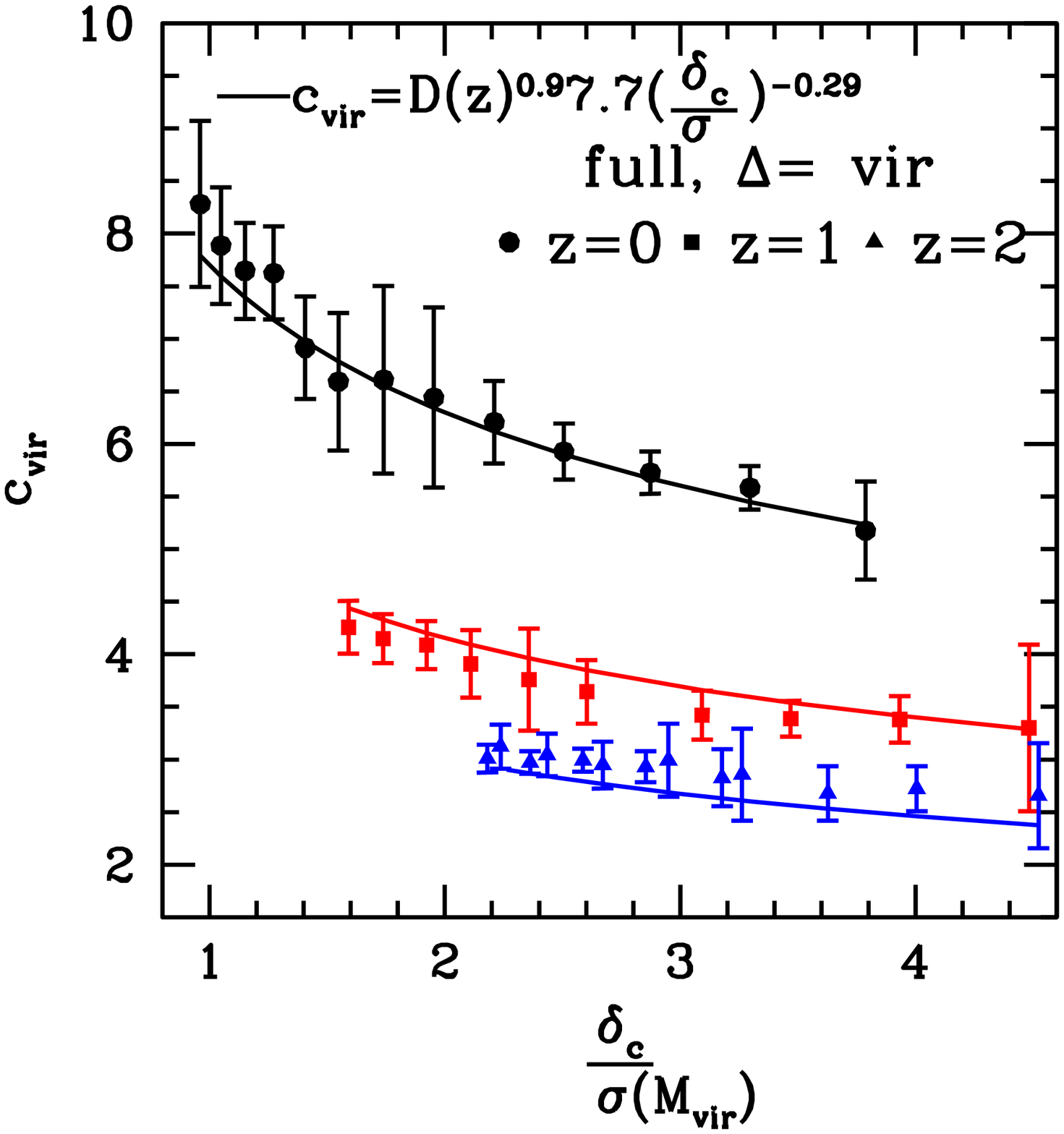}} 
        \resizebox{3.6in}{3.6in}{\includegraphics{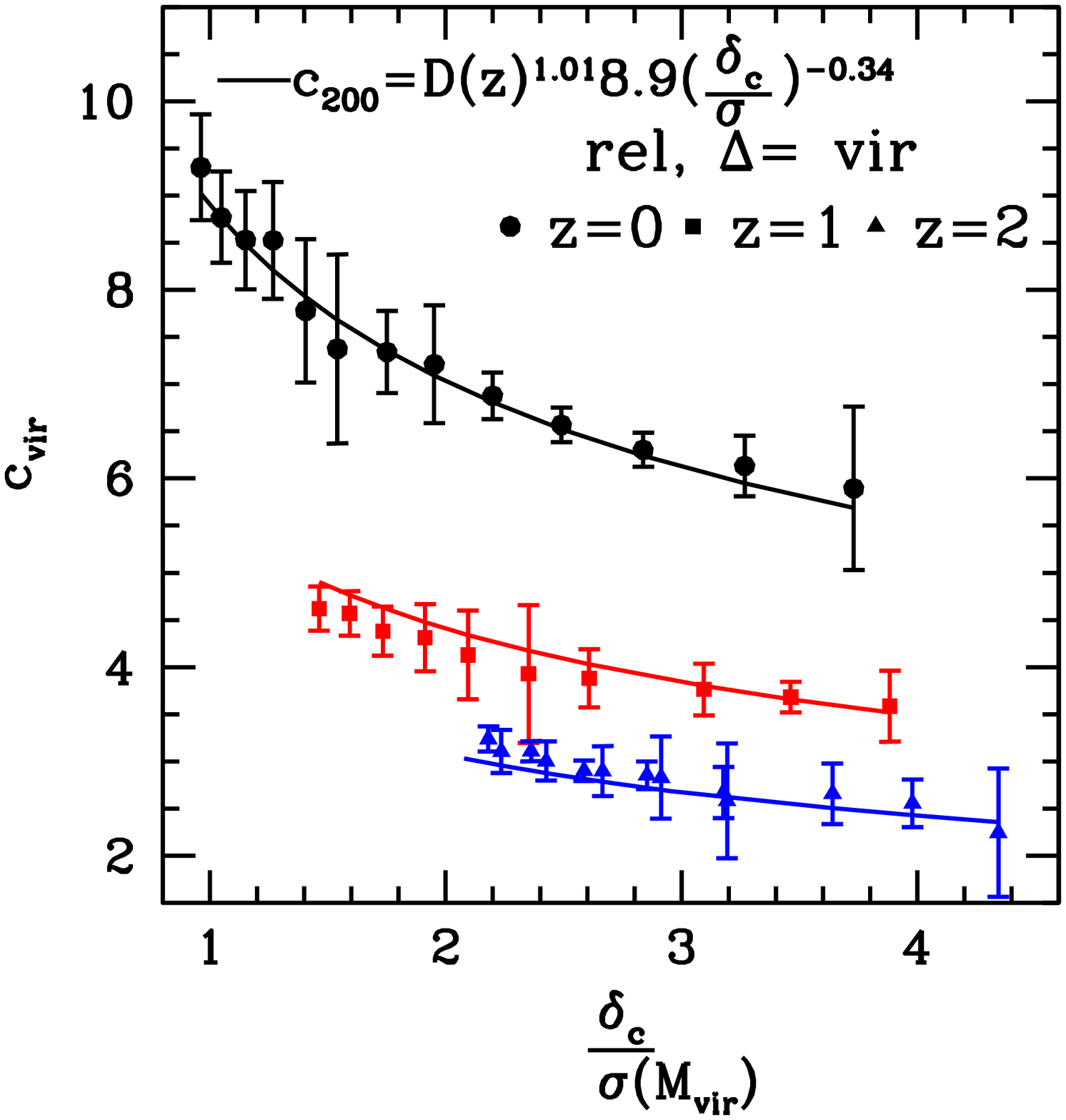}} 
        \end{tabular}
    \caption{$c-\nu$ relations at radii $r=R_{200}$ and $r=R_{vir}$
      for $z=0$, 1, and 2, for the relaxed and full samples where
      $\nu=\delta_c/\sigma(M_\Delta)$ with $\delta \sim$ 1.676, varying
      only mildly with redshift. The lines are global fits to the data
      points using a simple assumption for redshift evolution.   
    }
\label{fig:c-Mnu}
  \end{center}
\end{figure*}

In our simulations, we identify halos using a fast parallel
friends-of-friends (FOF) finder \citep{woodring11} with linking length
$b=0.2$. (The Appendix contains a discussion of how this choice can
affect results.) The effects of major sub-structure, relevant for
roughly a quarter of the halos (e.g., \citealt{lukic09}) is checked by
using morphological cuts mentioned later below. Since we are concerned
only with the mass profiles, and not the dynamical state of the halo,
we do not use any velocity information (for instance, whether to
unbind particles or not). Once a halo is found, we define its center
via a density maximum criteria -- the location of the particle with
the maximum number of neighbors. This definition of the halo center is
very close to that given by the potential minima. Given a halo center,
we grow spheres around it and compute the mass in radial bins. Note
that even though an FOF finder is used, the actual halo mass is
defined by a spherical overdensity method, consistent with what is
done in observations. (For discussions on halo mass, see
\citealt{white01}, \citealt{lukic09}, and \citealt{more11}.) Although
the mass could be measured independently of the concentration we fit
both together to the halo profile, as this is potentially less
sensitive to fitting bias. (In practice it makes little difference.)

We write the NFW profile as
\begin{equation}
\rho(r)= \frac{\delta\rho_{\rm{crit}}}{(r/r_s)(1+r/r_s)^2}
\label{eq:nfw}
\end{equation}
where $\delta$ is a characteristic dimensionless density, and $r_s$ is
the scale radius of the NFW profile. The concentration of a halo is
defined as $c_{\Delta}=r_{\Delta}/r_s$, where $\Delta$ is the
overdensity with respect to the {\it critical density} of the
Universe, $\rho_{\rm{crit}}=3H^2/8\pi G$, and $r_{\Delta}$ is the
radius at which the enclosed mass, $M_{\Delta}$, equals the volume of
the sphere times the density $\Delta \rho_{\rm{crit}}$. We compute
concentrations at two radii corresponding to $\Delta=200$ and
$\Delta=\Delta_{\rm{vir}}$, corresponding in turn to $c_{200}=
R_{200}/r_s$ and $c_{vir} = R_{vir}/r_s$. The value of $\Delta_{vir}$
is given by the spherical top-hat collapse model; it changes with
redshift and cosmology and, for $\Lambda$CDM, can be approximated by a
fitting formula $\Delta_{\rm{vir}}=18\pi^2+82x-39x^2$ with
$x=\Omega_m(z)-1$,
$\Omega_m(z)=\Omega_m(1+z)^3/(\Omega_m(1+z)^3+\Omega_\Lambda)$
\citep{bryan98}. For our reference cosmology, $\Delta_{vir}$ varies
from $\sim$ 95-170 over the range $z=0-2$. We also provide a fit for
the overdensity of $200$ times the {\it mean density}, $\rho_b$, of
the universe at a particular redshift, $z$. Written in terms of the
critical density, this corresponds to $\Delta=200\Omega_m(z)$ which
varies from $50-180$, over the range $z=0-2$ for our reference
cosmology.

The mass enclosed within a radius $r$ for an NFW halo profile is given
by
\begin{equation}
M(<r)= \frac{m(c_\Delta r/r_{\Delta})}{m(c_{\Delta})}M_{\Delta},
\label{eq:nfwmass}
\end{equation}
where $m(y)=\ln(1+y)-y/(1+y)$. 
The mass in a radial bin is then
\begin{equation}
M_i=M(<r_i)-M(<r_{i-1}).
\label{eq:nfwmassbin}
\end{equation}
We fit Eq.~\ref{eq:nfwmassbin} to the mass contained in the radial
bins of each halo, by minimizing the associated value of $\chi^2$ as
\begin{equation}
\chi^2= \sum_i \frac{(M_i^{sim}-M_i)^2}{(M_i^{sim})^2/n_i} 
\label{eq:fit}
\end{equation}
where the sum is over the radial bins, $n_i$ is the number of
particles in a radial bin, $M_i^{sim}$ is the mass in bin $i$
calculated from the simulations and $M_i$ is the mass calculated
assuming the NFW profile. The advantage of fitting mass in radial bins
rather than the density is that the bin center does not have to be
specified. Note that we explicitly account for the finite number of
particles in a bin. This leads to a slightly larger error in the
profile fitting but minimizes any possible bias due to the finite
number of particles, especially near the halo center.

We fit for two parameters -- the normalization of the profile and the
concentration. Halo profiles are fitted in the radial range of
approximately $(0.1-1)R_{vir}$. This choice is motivated partly by the
observations of concentrations that typically exclude the central
region of clusters (e.g., observations by \citealt{oguri11}, to which
we compare our results in Section~\ref{sec:comp}). More significantly,
however, this excludes the central core which is sensitive to the
effects of baryonic physics and numerical errors arising from
limitations in both mass and force resolution, as discussed in the
Appendix. As already mentioned, \cite{duffy10} have shown that, at
$r<0.1R_{vir}$, halo profiles are sensitive to the impact of baryons
with the profiles being affected at $r=0.05 R_{vir}$ by as much as a
factor of 2. In the Appendix we discuss the robustness of the obtained
$c-M$ relation as the fitting range is varied; we find that different
fitting ranges -- chosen with a fair degree of lattitude -- agree with
each other to better than $10\%$ accuracy (Figure~\ref{fig:cmsys}).

The $c-M$ relation is calculated  by weighing the individual
concentrations by the halo mass, 
\begin{equation}
c(M)= \frac{\sum_i c_i M_i}{\sum_i M_i}
\label{eq:mean_c}
\end{equation}
where the sum is over the number, $N_i$, of the halos in a mass
bin. The mass of the bin is given by 
\begin{equation}
M= \sum_i M_i/N_i.
\label{eq:m}
\end{equation}
The error on $c(M)$ is the mass-weighted error on the individual fits
plus the Poisson error due to the finite number of halos in an
individual bin added in quadrature,
\begin{equation}
\Delta c(M)= \sqrt {\left (\frac{\sum_i \Delta c_i M_i}{\sum_i
      M_i}\right )^2+ \frac{c^2(M)}{N_i}} ,
\label{eq:errmean}
\end{equation} 
where $\Delta c_i$ is the individual concentration error for each
halo. The first term dominates towards the lower mass end where the
individual halos have smaller number of particles and the second term
dominates towards the higher mass end, where there are fewer halos to
average over.

Figure~\ref{fig:c-M} shows the mean $c-M$ relation obtained from our
simulation runs weighted by the mass (Eq.~\ref{eq:mean_c}). We show
the $c-M$ relation both for relaxed halos and for the full (relaxed +
non-relaxed) sample.

To select the relaxed sample we use criteria similar to those of
\cite{neto07} and \cite{duffy08}, defining relaxed halos as those in
which the difference between the location of the center-of-mass and
the center density maximum is $< 0.07 R_{vir}$ (see also
\citealt{thomas01}). \cite{neto07} have used two additional criteria
to select their relaxed sample but found that the difference in the
center of halos method already selected most of the relaxed sample. We
do not impose their additional criteria as it would lead to
insignificant changes in our sample selection. At $z=0$, the relaxed
fraction varies from $0.73-0.6$ from $M_{200}=10^{12}
-7.5\times10^{14} \mau$, and the results for this fraction are
consistent with those found by \cite{neto07}. As the redshift
increases, one would expect this ratio to decrease as a function of
mass. For the bins centered at $M_{200}=2.47\times10^{12} \mau$, and
$M_{200}=1.39\times10^{13} \mau$, the values are $0.77$, $0.69$,
$0.67$, and $0.74$, $0.63$, $0.63$, at $z=0$, $1$, $2$
respectively. At $M_{200}=1.39\times10^{14} \mau$, the values are
$0.63$, $0.48$, at $z=0$, $1$ (insufficient statistics at $z=2$).

From Fig.~\ref{fig:c-M} it is clear that the $c-M$ relation becomes
considerably flatter at $z>0$, with the full sample relation
flattening more at higher redshift compared to that for the relaxed
sample. The $c-M$ relation for the relaxed sample has on an average a
10\% higher amplitude compared to that for the full sample. The $c-M$
relation at the radius corresponding to $\Delta=\Delta_{vir}$ has
about a $30\%$ higher amplitude compared to that at $\Delta=200$.

Because the cosmologies considered are essentially the same, we can
directly compare our results with those of \cite{duffy08}, although
their statistics become somewhat limited near the upper end of halo
masses. We find that at $z=0$, at cluster mass scales, their $c-M$
amplitude is about 15\% lower compared to our results. At $z=2$, the
results from \cite{duffy08} are about 15\% higher, but with
significant scatter. In general, their redshift evolution appears to
be slightly compressed, more so in the case of relaxed halos. Within
the statistical limitations mentioned, we may consider the comparison
to be quite reasonable. At $z=0$, our results can be fitted very
accurately by a power law with the exponent, $\alpha=-0.08$ and
$-0.084$ for the full and the relaxed sample. The logarithmic slope
corresponding to the full sample is precisely that found by
\cite{hayashi07} using the halo-density cross-correlation applied to
data from the MS. The normalization, however, is not expected to be
the same because of the high value of $\sigma_8=0.9$ chosen for the
MS. Note that three different analyses of the MS (and an associated
smaller-volume, higher mass resolution run) have produced slightly
discrepant results, differing from each other at the $10-20\%$ level
\citep{neto07, gao07, hayashi07}. This is probably a useful empirical
measure of the systematic issues inherent to halo selection and
fitting. In general, the MS results are consistently higher at all
redshifts by about $15\% $ at $z=0$ to about $30\%$ at $z=2$ largely
because of the higher value of $\sigma_8$.

\subsection{The $c-\nu$ relation}
\label{subsec:c_nu}
We find that the $c-M$ relation becomes almost flat at higher
redshift, with $c_{200}\sim 3$ in a way similar to the findings of
\cite{gao07} (although for a higher $\sigma_8$ in their
case). \cite{gao07} use the Einasto profile for stacked halos -- with
its extra shape parameter -- to account for the flatter $c-M$ relation
at higher $z$. Alternatively, we ask if it is possible to explain the
flattening of the $c-M$ relation with redshift using only the NFW
profile without adding an extra parameter. To do this, we follow the
strategy adopted in parameterizing halo mass functions and investigate
the concentration measurements as a function of the rms density
fluctuation $\sigma(M,z)$, rather than $M$ (for each of the three halo
mass definitions). As the central variable, we use the peak height
parameter, $\nu=\delta_c(z)/\sigma(M,z)$, where $\delta_c(z)$ is the
linear collapse threshold. ($\delta_c=1.673$ for the reference
cosmology and varies only mildly with cosmology and redshift.)
$\sigma(M,z)$ specifies the variance of the matter fluctuations over
the scale $\propto M^{1/3}$ at a redshift $z$. As shown in
Figure~\ref{fig:c-Mnu}, the shape of the $c-\nu$ relation is
approximately constant over the redshift range $z=0-2$, in contrast to
the shape of the $c-M$ relation.

Overall, the evolution of $c_{200}(\nu)$ proceeds as $\sim D(z)^{0.5}$
where $D(z)$ is the linear growth factor at redshift $z$, or by about
$30\%$ from $z=0-2$.  The overall $z$-evolution in our work is roughly
consistent with the $z$-evolution seen in the MS result of
\cite{gao07}. The evolution of $c_{vir}(\nu)$ possesses a somewhat
larger dynamic range and the evolution goes as $\sim D(z)$. The slope
of the $c-\nu$ relation is slightly larger for $\Delta=200$ compared
to that for $\Delta=\Delta_{vir}$. The amplitude of the $c-\nu$
relation is only a little larger for the relaxed sample compared to the
full sample (by about $\sim 10\%$).

Fitting formulae for $c(\nu)$ as derived from the simulations for the
reference $\Lambda$CDM cosmology are given in Table~\ref{tab:fit}. We
also provide an approximate fitting formula relating $\nu$ and $M$
valid for all overdensities, redshift, and cosmology, which can then
be used to convert the relations for $c-\nu$ to those for
$c-M$. Table~\ref{tab:fit} also provides the $c-\nu$ relation for
$\Delta=200\Omega_m(z)$ corresponding to an overdensity of
$200\rho_b$. At $z=0$, for the reference cosmology used here,
$200\rho_b$ is about half of the virial overdensity. Consequently, the
$c-\nu$ amplitude is about $30\%$ higher compared to the amplitude at
the virial density. At $z=1$ and 2, the mean density and the virial
density become comparable. Thus the $c-\nu$ relation for
$\Delta=200\Omega_m(z)$ has more $z$-evolution when compared to that
for the virial density.

\begin{table}
\begin{center} 
\caption{$c(\nu)-\nu$ fitting formulae.}
\label{tab:fit}
\begin{tabular}{c|c|c|c}
\hline
  & $\Delta=200$ & $\Delta=\Delta_{vir}$ & $\Delta=200 \rho_b$\\
\hline\hline
 & & & \\
full & $D(z)^{0.54}5.9\nu^{-0.35}$ & $D(z)^{0.9}7.7\nu^{-0.29}$ &
$D(z)^{1.15}9.0\nu^{-0.29}$\\  
 & &   & \\
relaxed & $D(z)^{0.53}6.6\nu^{-0.41}$ & $D(z)^{1.01} 8.9\nu^{-0.34}$ &
$D(z)^{1.2} 10.1\nu^{-0.34}$\\ 
\hline\hline
\end{tabular}\\
\begin{tabular}{l|l}
Std. Dev. &$\sigma_c= 0.33c_{\Delta}$\\
 & \\
$\nu-M$ &$\nu(M,z) \approx
\frac{1}{D(z)}\left[1.12\left(\frac{M}{5\times 10^{13} \mau} \right
  )^{0.3}+0.53\right]$\\\\
\hline\hline
\end{tabular}
\end{center}
\end{table}

\subsection{The distribution of concentrations}
\label{sub:concdist}

The mean $c-M$ relation needs to be augmented with a good quantitative
understanding of the concentration scatter around the mean, especially
at cluster-scale masses, where simulations have historically suffered
from lack of volume coverage. Predictions for the distribution of the
concentration are particularly valuable since they can be used to
check for selection biases in observations. As an example, if there is
a concern that lensing-based searches are likely to be biased towards
high concentration halos, then, at a given mass bin, one can test for
this bias by comparing to the predicted theoretical distribution. For
this method to work, there should be enough objects at a given mass, a
target that will be attained in the near future. 

\begin{figure}
  \begin{center}
    \begin{tabular}{c}
          \resizebox{3.6in}{3.6in}{\includegraphics{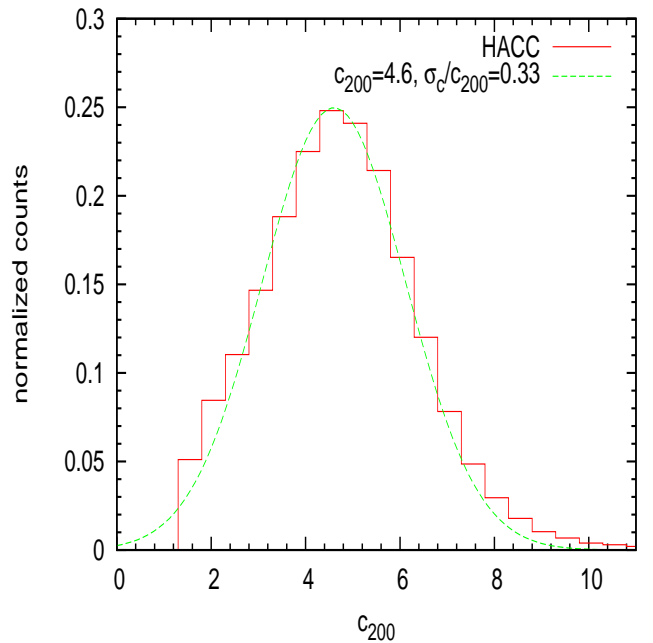}}   
    \end{tabular}
          \caption{$c-M$ distribution at a mass bin centered at
            $1.5\times 10^{14}$~$h^{-1}$M$_{\odot}$ using results from the
            HACC simulation at $z=0$. Lines show a Gaussian
            distribution with standard deviation $\sigma_c/c \sim 0.33$.}
\label{fig:cdist_gauss}
  \end{center}
\end{figure}

We have computed the concentration distribution for a large set of
cosmologies, a subset of which we discuss here. Previous studies
\citep{jing00, shaw06, neto07, duffy08}, have fitted the concentration
distribution to a log-normal distribution. However, this distribution
is also very well described by a Gaussian as noted by \cite{lukic09}
and \cite{reed11}. We have found that a Gaussian distribution provides
a very good fit to our data, with relatively small non-Gaussian
tails. As a representative example, we show the distribution of
$c_{200}$ (full halo sample at $z=0$) for the mass bin centered at
$1.5\times 10^{14} h^{-1}M_\odot$ in Figure~\ref{fig:cdist_gauss}.

Assuming a Gaussian distribution, the standard deviation in the $c-M$
relation can be calculated as,
\begin{equation}
\sigma_c(M)= \sqrt{\frac{\sum_i c_i^2 M_i}{\sum_i M_i}- c(M)^2}
\end{equation}
and the associated error in determining the scatter is the Poisson
error in each bin,
\begin{equation}
\Delta \sum_c(M)= \sum_c(M)/\sqrt{N_i},
\end{equation}
where $N_i$ is the number of halos in the mass bin with mass $M$.

\begin{figure}
  \begin{center}
    \begin{tabular}{c}
      \resizebox{3.6in}{3.6in}{\includegraphics{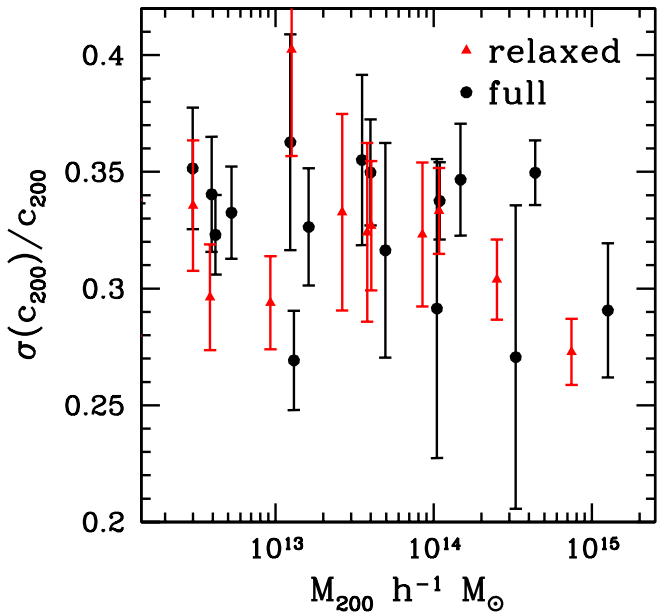}}
            \end{tabular}
            \caption{The $c_{200}-M_{200}$ distribution for the
              relaxed and full halo samples characterized by the ratio
              of the standard deviation to the mean value of
              $c_{200}$. All three redshifts are plotted. Note that
              $\sigma_c/c$ shows no redshift evolution. The case of
              $\Delta=\Delta_{vir}$ shows identical behavior.}
\label{fig:cMsigma}
  \end{center}
\end{figure}

As illustrated in Figure~\ref{fig:cdist_gauss} for the case of halos
in the mass bin at $1.5\times10^{14}$~$h^{-1}$M$_{\odot}$, at $z=0$,
the standard deviation of the Gaussian distribution is roughly
$\sigma_c= 0.33 c$, over our mass and redshift range, with mild
dependence on the mass at the very high mass end, such that for
$M_{200} > 8\times 10^{14} \mau$, $\sigma_c \sim 0.28 c$
(Figure~\ref{fig:cMsigma}). These results are in very good agreement
with \cite{reed11} who find $\sigma_c \sim 0.28 c$ for an analysis of
halos extracted from the MS. If instead, for comparison purposes, we
fit our concentration distribution using the log-normal function, we
get $\sigma(\log_{10}(c_{200}))=0.16$ for the full sample and $0.12$
for the relaxed sample. This may be compared to the result of
\cite{duffy08}, who obtain $0.15$ and $0.11$, respectively, and to
\cite{neto07} who find $0.14$ and $0.1$. Our scatter is therefore
about $5-10\%$ higher than the results of \cite{duffy08} and
\cite{neto07}. As shown by \cite{neto07}, the variance is at a minimum
for the radial range $(0.05-1)R_{vir}$. As noted earlier, we fit the
halo profile over the range $(0.1-1)R_{vir}$ which may account for the
$\sim 7\%$ larger value of the standard deviation. Our choice trades
off a slight increase in scatter for robustness against systematic
effects, as previously discussed.  Figure~\ref{fig:cMsigma} shows that
the relation $\sigma_c=0.33 c$ is more or less independent of mass,
redshift, or the dynamical state of the halo. The relation also
remains constant when the cosmology is varied
(Figure~\ref{fig:sigcwcdm}). This means that the standard deviation of
the concentration distribution depends on cosmology, redshift, or the
dynamical state, in the same way as the mean concentration, confirming
the initial finding of \cite{dolag04} from a small sample of simulated
halos, but spanning multiple dark energy cosmologies.

\section{$c$-M relation for $w$CDM cosmologies}
\label{sec:wcdm}

\begin{table*}[t]
\begin{center} 
\caption{\label{table_wcdm} Parameters for the 18 cosmological models
  used to study the $c-M$ relation}      
\hspace{0cm}
\begin{tabular}{ccccccccc}
\tableline\tableline
\#& $\omega_m$ & $\omega_b$ & $n_s$ & $-w$ & $\sigma_8$ & $h$ & M
&variation\\ 
 & & & & & & & $10^{14} M_\odot$& \%\\
\hline 
1 & 0.1539 & 0.0231 & 0.9468 & 0.816 & 0.8161  & 0.5977 & 13.3 & 0\\
3 & 0.1324 & 0.0235 & 0.9984 & 0.874 & 0.8484 & 0.6763 & 9.96  & +15\\
4 & 0.1381 & 0.0227 & 0.9339 & 1.087 & 0.7000 & 0.7204 & 4.42& -18\\
5 & 0.1358 & 0.0216 & 0.9726 & 1.242 & 0.8226 & 0.7669 & 7.20 & -20\\
7 & 0.1268 & 0.0223 & 0.9210 & 0.700 & 0.7474 & 0.6189 & 7.30 & +15\\
8 & 0.1448 & 0.0223 & 0.9855 & 1.203 & 0.8090 & 0.7218 & 8.04& -15\\
9 & 0.1392 & 0.0234 & 0.9790 & 0.739 & 0.6692 & 0.6127 & 4.98& +10\\
12 & 0.1223 & 0.0225 & 1.0048 & 0.971 & 0.6271 & 0.7396 & 2.26 & +14\\
13 & 0.1482 & 0.0221 & 0.9597 & 0.855 & 0.6508 & 0.6107 & 4.78 & +5\\
14 & 0.1471 & 0.0233 & 1.0306 & 1.010 & 0.7075 & 0.6688 & 5.42 & 0\\
15 & 0.1415 & 0.0230 & 1.0177 & 1.281 & 0.7692 & 0.7737 & 5.47 & -15\\
16 & 0.1245 & 0.0218 & 0.9403 & 1.145 & 0.7437 & 0.7929 & 4.22 & -10\\
17 & 0.1426 & 0.0215 & 0.9274 & 0.893 & 0.6865 & 0.6305 & 5.50 & 0\\
18 & 0.1313 & 0.0216 & 0.8887 & 1.029 & 0.6440 & 0.7136 & 3.05 & -10\\
19 & 0.1279 & 0.0232 & 0.8629 & 1.184 & 0.6159 & 0.8120 & 1.88 & -15\\ 
20 & 0.1290 & 0.0220 & 1.0242 & 0.797 & 0.7972 & 0.6442 & 8.24 & +20\\ 
30 & 0.1234 & 0.0230 & 0.8758 & 0.777 & 0.6739 & 0.6626 & 4.09 & +10\\ 
37 & 0.1495 & 0.0228 & 1.0233 & 1.294 & 0.9000 & 0.7313 & 11.7 & -15\\
\tableline\tableline
\vspace{-1.5cm}
\tablecomments{ The second column from the right shows the mass
  corresponding to $\nu=3$ for each cosmology at $z=0$. The right-most
  column shows the approximate variation of the mean $c-M$ relation
  with respect to the reference run.} 
\end{tabular}
\end{center}
\end{table*}

\begin{figure*}
  \begin{center}
    \begin{tabular}{ccc}
      \resizebox{2.5in}{2.5in}{\includegraphics{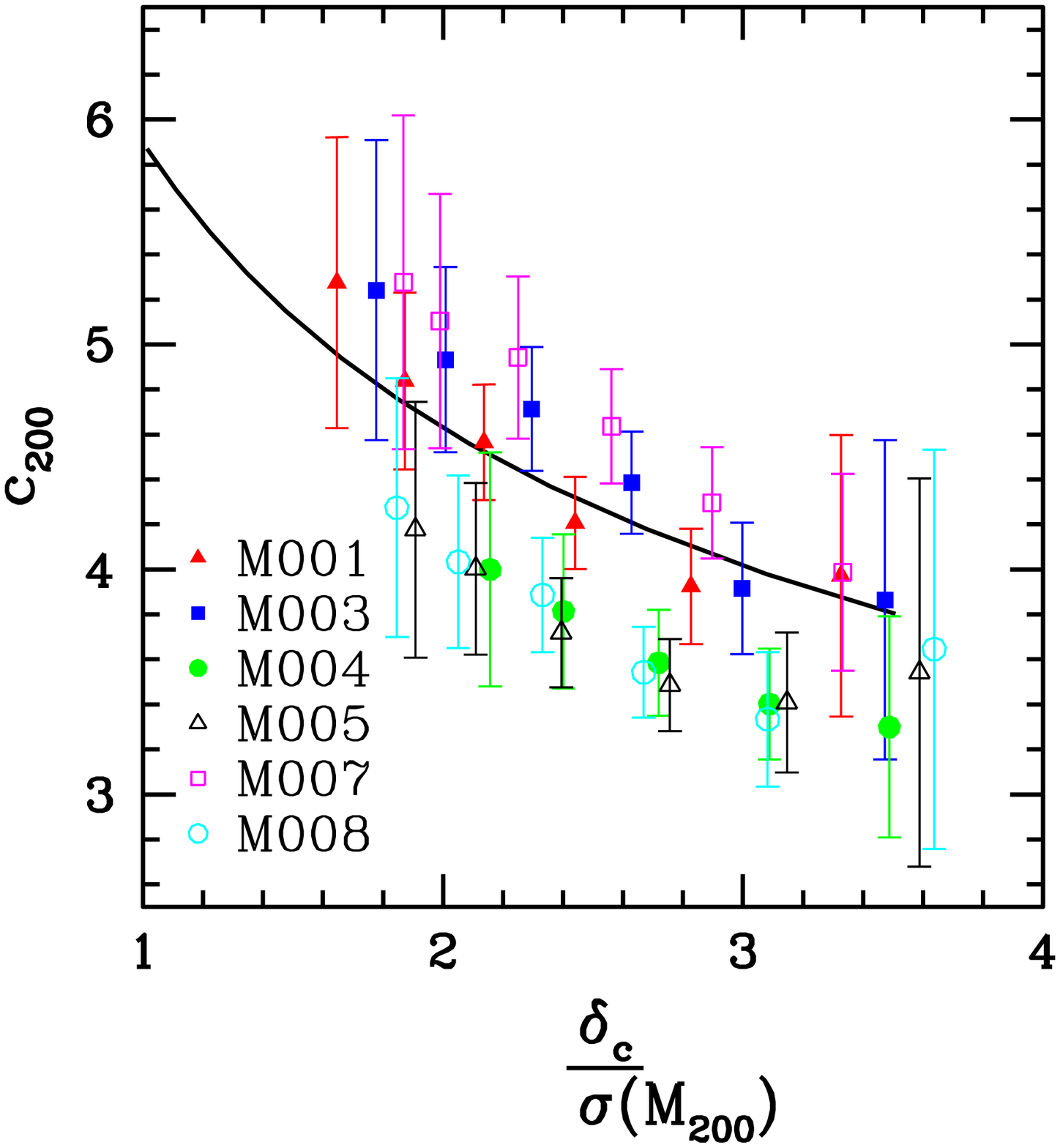}}
       \resizebox{2.5in}{2.5in}{\includegraphics{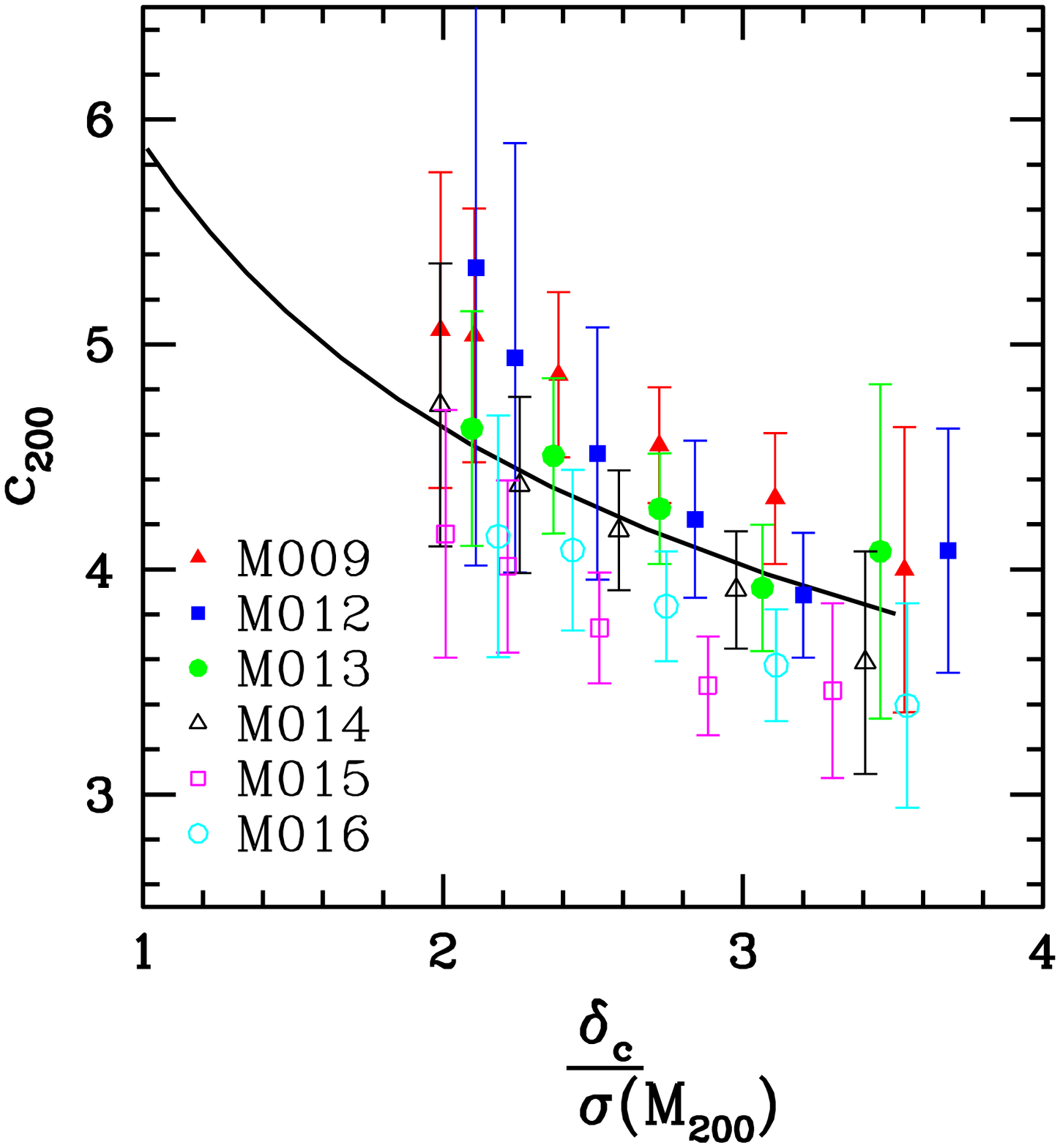}}
             \resizebox{2.5in}{2.5in}{\includegraphics{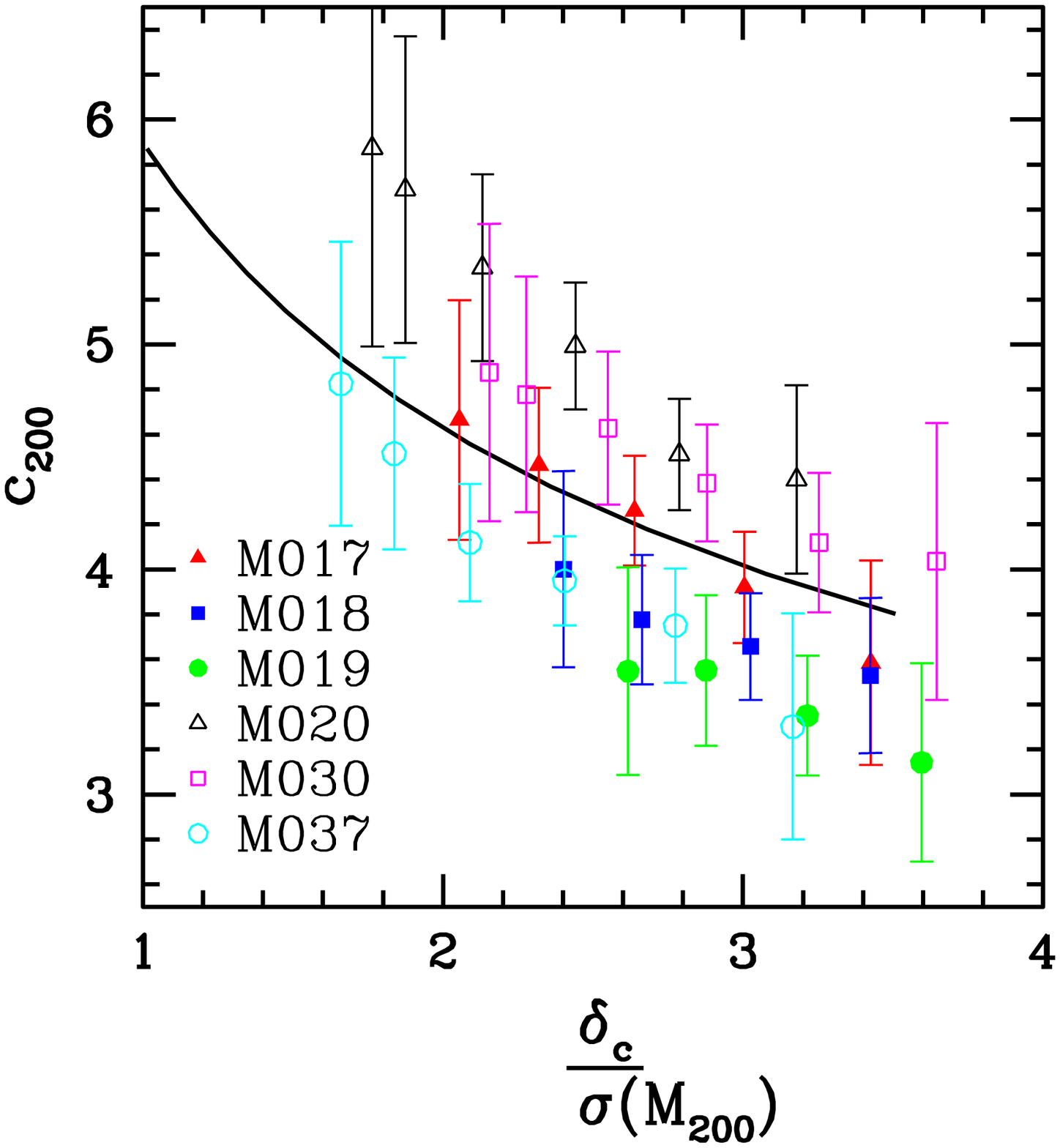}}
             \end{tabular}
             \caption{Mean $c-M$ relation at $z=0$ (full sample) when
               $w$CDM parameters are varied. The solid curve in all
               three panels is the fit to the reference $\Lambda$CDM
               cosmology as specified in Table~\ref{tab:fit}.}
\label{fig:meancwcdm}
  \end{center}
\end{figure*}

\begin{figure}
  \begin{center}
    \begin{tabular}{c}
         \resizebox{3.65in}{3.4in}{\includegraphics{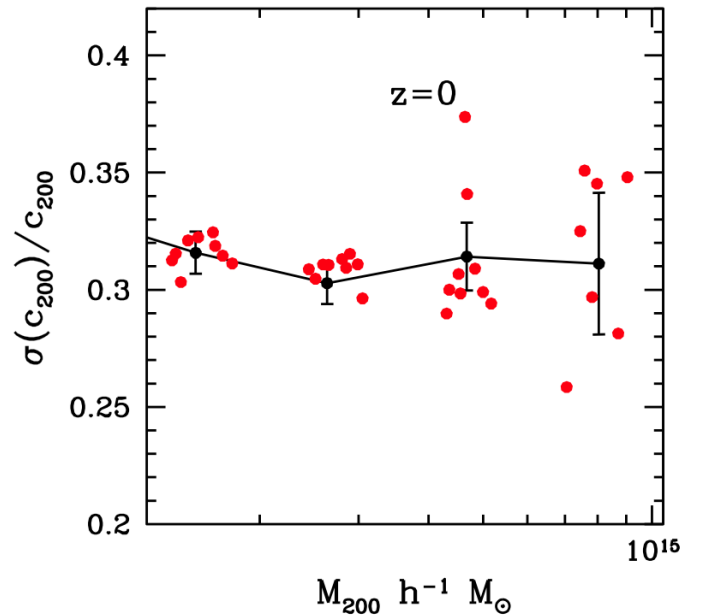}}
        \end{tabular}
        \caption{The $c_{200}-M_{200}$ distribution at $z=0$ (full
          sample) when $w$CDM parameters are varied, following the
          characterization of Fig.~\ref{fig:cMsigma}. The scatter is
          larger at high masses due to lower numbers of halos in
          high-mass bins.}
\label{fig:sigcwcdm}
  \end{center}
\end{figure}

In this section we study how the halo profiles, and hence the
concentration, vary with cosmology. We use 18 different runs, each
with a volume of $(1.3~\rm{Gpc})^3$ and $1024^3$ particles. The runs
are carried out using {\sc GADGET-2} and each run has a different set
of $w$CDM parameters. These simulations are a subset of the Coyote
Universe suite (see \citealt{heitmann09} and \citealt{heitmann10} for
details). The simulation suite consists of 38 runs covering the
$2\sigma$ range of $w$CDM parameter space as constrained by WMAP 5
year results \citep{wmap5}. We choose 18 runs (plus the reference
cosmology run) out of the 38 to show the cosmology dependence and
retain the model numbering from the original Coyote runs. The runs
have a coarser force resolution than the GS and HACC simulations. The
effect of this is considered in the Appendix, where it is shown that
the G run has a systematically lower concentration compared to the
HACC run over the same mass scale by about 5-10\%
(Figure~\ref{fig:cmsys}, left panel). To compensate for this minor
underestimate, we rescale concentrations obtained from the $w$CDM runs
by a factor of $1.05$, checking for correctness by comparing against
the fit obtained for the reference cosmology.

Figure~\ref{fig:meancwcdm} shows the variation of the $c-\nu$ relation
with respect to the best-fit WMAP5 cosmology. The mean $c-\nu$
relation varies by about $\pm$ 20\% over the currently allowed $w$CDM
cosmological parameter range. Note that $\nu$ already accounts for
some of the cosmology dependence of the $c-M$ relation, so a part of
the variation is actually hidden. Since we have already found that
expressing $c$ as a function of $\nu$ explains the redshift evolution
of NFW halo profiles, we illustrate the cosmology dependence using the
$c-\nu$ relation in place of the $c-M$ relation.

Table~\ref{table_wcdm} shows the approximate difference between the
(corrected) mean $c-M$ relation seen in each of the $w$CDM runs
compared to the mean $c-M$ relation obtained for the reference
$\Lambda$CDM cosmology. Note that although most of the variation in
the $c-M$ relation is in the overall amplitude, the slope also changes
for some of the models (e.g., M003, M012). Interestingly, we find that
some of the models show no variation compared to the reference,
although these models differ across the range of cosmological
parameters. For example, M014 and M017 both have lower $\sigma_8$
compared to the reference model, but show essentially no variation --
parameters other than $\sigma_8$ are clearly also active. The standard
deviation of the concentration distribution, on the other hand,
changes in the same way as the mean, leaving the ratio $\sigma_c/c$
almost universal \citep{dolag04}. Figure~\ref{fig:sigcwcdm} shows that
the $\sigma_c/c$ varies by $< 5\%$ over the range of $w$CDM
cosmologies.

Semianalytical `toy models' based on Press-Schechter arguments
\citep{ps74} have attempted to model the cosmology and redshift
dependence of halo concentrations via the underlying dependence on the
matter power spectrum and the evolution history of the universe
\citep{nfw2, eke01, bullock99}. The model of \cite{nfw2} has two free
parameters. It defines the halo formation redshift by requiring that
half of the final halo mass be in progenitors with masses of some
fraction of the final mass. The characteristic density scale of the
NFW profile is then set by assuming it to be some multiple of the
cosmic density at the redshift of halo formation. The mass fraction
and the density multipliers are given by fitting to
simulations. \cite{bullock99} modified this prescription by redefining
the formation redshift as the redshift where the nonlinear mass scale
$M_*$ is some fraction of the final halo mass. They predicted the
concentration as a multiple of the ratio of scale factors at the
formation and collapse redshifts. Again, the fraction and multiplier
are floating parameters, determined by fitting. Finally, \cite{eke01}
used a single parameter (calibrated using simulations) to connect the
collapse redshift with the effective amplitude of the power spectrum
at cluster scales.

Even though the models reproduce the expected behavior of the $c-M$
relation discussed in the Introduction, quantitatively their success
has been decidedly mixed -- results have been satisfactory over
limited dynamic ranges when fitted to simulations, but tended to break
down as the range is extended or cosmological parameters are
varied. Particularly significant for us is the breakdown at large halo
masses, characteristic of massive clusters at $z=0$ \citep{neto07,
  gao07, duffy08, zhao09}, with a corresponding breakdown at lower
masses, but at higher redshifts. Additionally, the redshift evolution
of the $c-M$ relation in the models is significantly stronger than is
actually seen in simulations.

Although we have not attempted to optimize model parameters, the
\cite{eke01} prescription agrees with our simulation results at
$10-20\%$ accuracy for $c_{200}$ for the $\Lambda$CDM cosmology (see
Table~\ref{tab:fit} for the fit). We find that the prescriptions of
\cite{dolag04} (using a growth factor ratio multilpier) do not explain
the cosmology variation in concentrations that we observe. For
instance, in our case, the growth factor only varies by $< 5\%$ over
the range of simulations used, whereas the variations of
concentrations seen is $\sim \pm$ 20\%.

Regarding the modeling of the scatter in the concentration, it is
natural to examine this in the context of different assembly histories
for halos with the same mass \citep{wechsler01, zhao03} (See also,
\citealt{cohn05}.) However, in their MS analysis, \cite{neto07} find
that the concentration scatter cannot be accounted for by differences
in the time of formation alone. Additional consequences of
environmental effects \citep{wechsler05} appear to be important
primarily for low-mass halos. Therefore one is driven to the general
conclusion that there is still no replacement for large-scale
simulations if reliable predictions for halo concentrations and the
distribution of concentrations are required.

\section{Comparison with observations}
\label{sec:comp}
\begin{table*}
\begin{center} 
\caption{observation data used in this paper}
\begin{tabular}{c|c|c|c|c}
\hline
 observation & method & rel./all & \# clusters & redshift range\\
\hline\hline
Oguri et. al. & Strong+Weak lensing& all & 28 & $0.28<z<0.64$\\
Okabe et. al. & Weak lensing & all & 30 & $0.15<z<0.3$\\
Wojtak \& Lokas & Kinematics & rel.& 41 & $z<$0.1\\
Vikhlinin et al & X-ray & rel. & 19 & $z<$0.2\\
Schmidt \& Allen & X-ray & rel. & 34 & $0.06<z<0.7$\\
Buote et. al. & X-ray & rel. & 26 & $z<0.23$\\
Ettori el. al. & X-ray & all & 44 & $0.1<z<0.3$\\
\hline\hline
\tablecomments{We use only objects with mass $>5\times 10^{13} \mau$
  from Buote et al.} 
\end{tabular}
\end{center}
\label{tab:obs}
\end{table*}

\begin{table*}
\begin{center} 
\caption{\label{table_vikhlinin} Updated masses and concentrations
  from the {\em Chandra} Cluster Cosmology Project}      
\hspace{0cm}
\begin{tabular}{ccccccccc}
\tableline\tableline
Cluster & M$_{500}$ (M$_{\odot}$) & $\delta$M (M$_{\odot}$) & $c_{500}$ & $+\delta_c$ & $-\delta_c$ & z \\ 
\hline 
a133 & 3.166$\times 10^{14}$ & 3.776$\times 10^{13}$ & 3.15 & 0.29 & 0.28  & 0.0569 \\
a262 & 8.310$\times 10^{13}$ & 7.272$\times 10^{12}$ & 3.48 & 0.30 & 0.30  & 0.0162 \\
a383 & 3.049$\times 10^{14}$ & 3.100$\times 10^{13}$  & 4.31 & 0.42 & 0.40  & 0.1883 \\
a478 & 7.668$\times 10^{14}$ & 1.010$\times 10^{14}$  & 3.57 & 0.27 & 0.26  & 0.0881 \\
a907 & 4.623$\times 10^{14}$ & 3.790$\times 10^{13}$  & 3.46 & 0.42 & 0.42  & 0.1603 \\
a1413 & 7.569$\times 10^{14}$ & 7.550$\times 10^{13}$  & 2.93 & 0.18 & 0.17 & 0.1429 \\
a1795 & 6.009$\times 10^{14}$ & 5.134$\times 10^{13}$  & 3.21 & 0.18 & 0.18  & 0.0622 \\
a1991 & 1.235$\times 10^{14}$ & 1.654$\times 10^{13}$  & 4.31 & 0.34 & 0.34  & 0.0592 \\
a2029 & 8.147$\times 10^{14}$ & 7.674$\times 10^{13}$  & 4.04 & 0.21 & 0.21  & 0.0779 \\
a2390 & 1.077$\times 10^{14}$ & 1.092$\times 10^{14}$  & 1.66 & 0.13 & 0.13  & 0.2302 \\
cl1159 & 1.056$\times 10^{14}$ & 2.578$\times 10^{13}$  & 1.77 & 0.38 & 0.24  & 0.0810 \\
MKW4 & 7.734$\times 10^{13}$ & 1.032$\times 10^{13}$  & 2.54 & 0.16 & 0.14 & 0.0199 \\
a2717 & 1.478$\times 10^{14}$ & 2.134$\times 10^{13}$  & 2.69 & 0.19 & 0.19 & 0.0498 \\
a3112 & 3.448$\times 10^{14}$ & 3.097$\times 10^{13}$  & 4.47 & 0.28 & 0.27 & 0.0761 \\
a1835 &1.245$\times 10^{15}$ & 1.342$\times 10^{14}$  & 2.81 & 0.17 & 0.17 & 0.2520 \\
a1650 & 4.683$\times 10^{14}$ & 1.736$\times 10^{13}$  & 3.74 & 0.19 & 0.19 & 0.0846 \\
a2107 & 2.361$\times 10^{14}$ & 3.928$\times 10^{13}$  & 3.38 & 0.28 & 0.25  & 0.0418 \\
a4059 & 3.496$\times 10^{14}$ & 2.691$\times 10^{13}$  & 2.95 & 0.09 & 0.09  & 0.0491 \\
rxj1504 & 1.068$\times 10^{15}$ & 1.768$\times 10^{14}$  & 3.16 & 0.38 & 0.38 & 0.2169 \\
\tableline\tableline
\vspace{-1.5cm}
\tablecomments{$\delta$M is the estimated error in the mass,
  $+\delta_c$ and $-\delta_c$, are the upper and lower error bounds
  for the concentrations. The masses are for $h=0.72$. } 
\end{tabular}
\end{center}
\end{table*}

\begin{figure*}
  \begin{center}
    \begin{tabular}{cc}
        \resizebox{3.6in}{3.6in}{\includegraphics{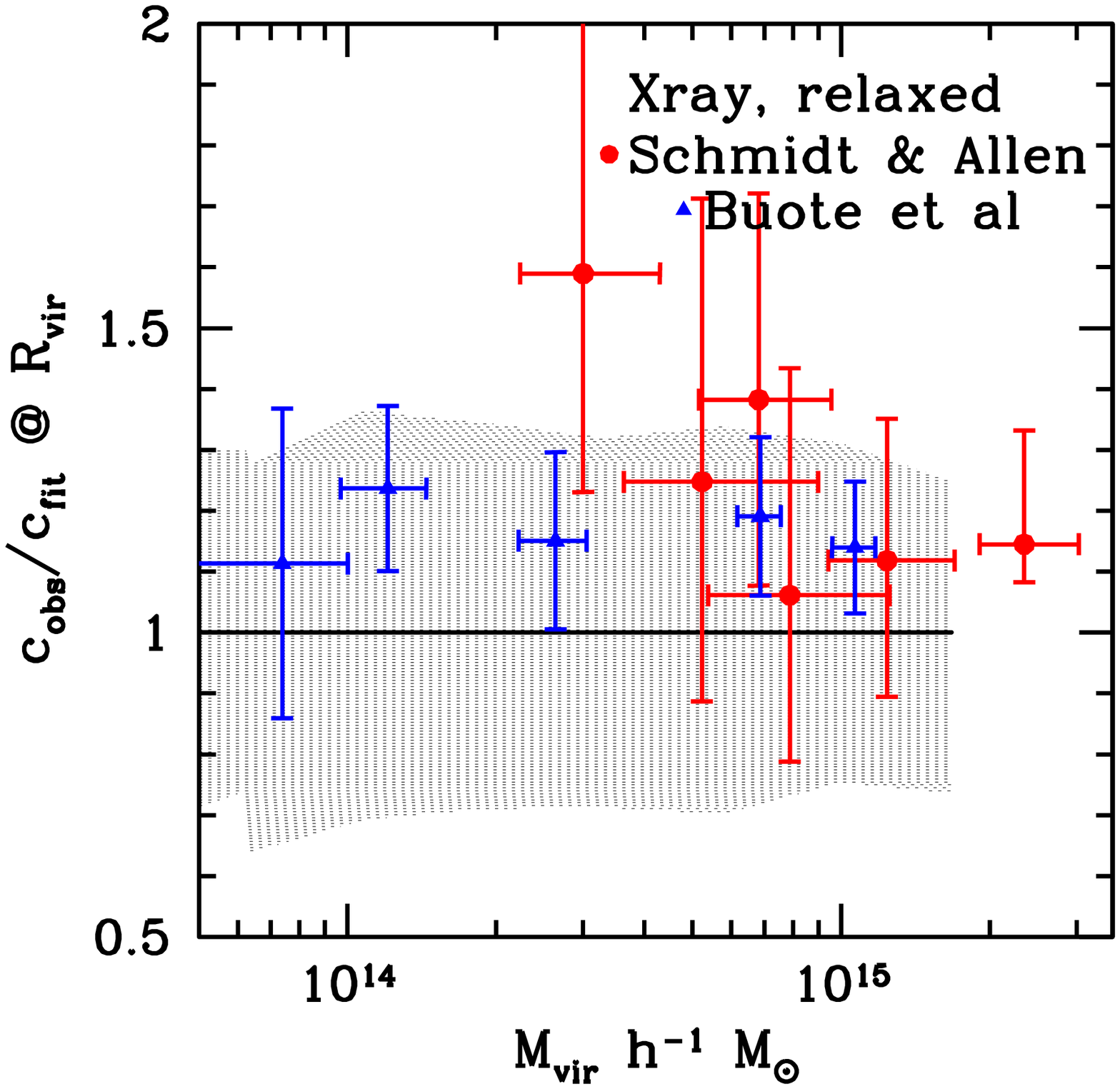}} 
        \resizebox{3.6in}{3.6in}{\includegraphics{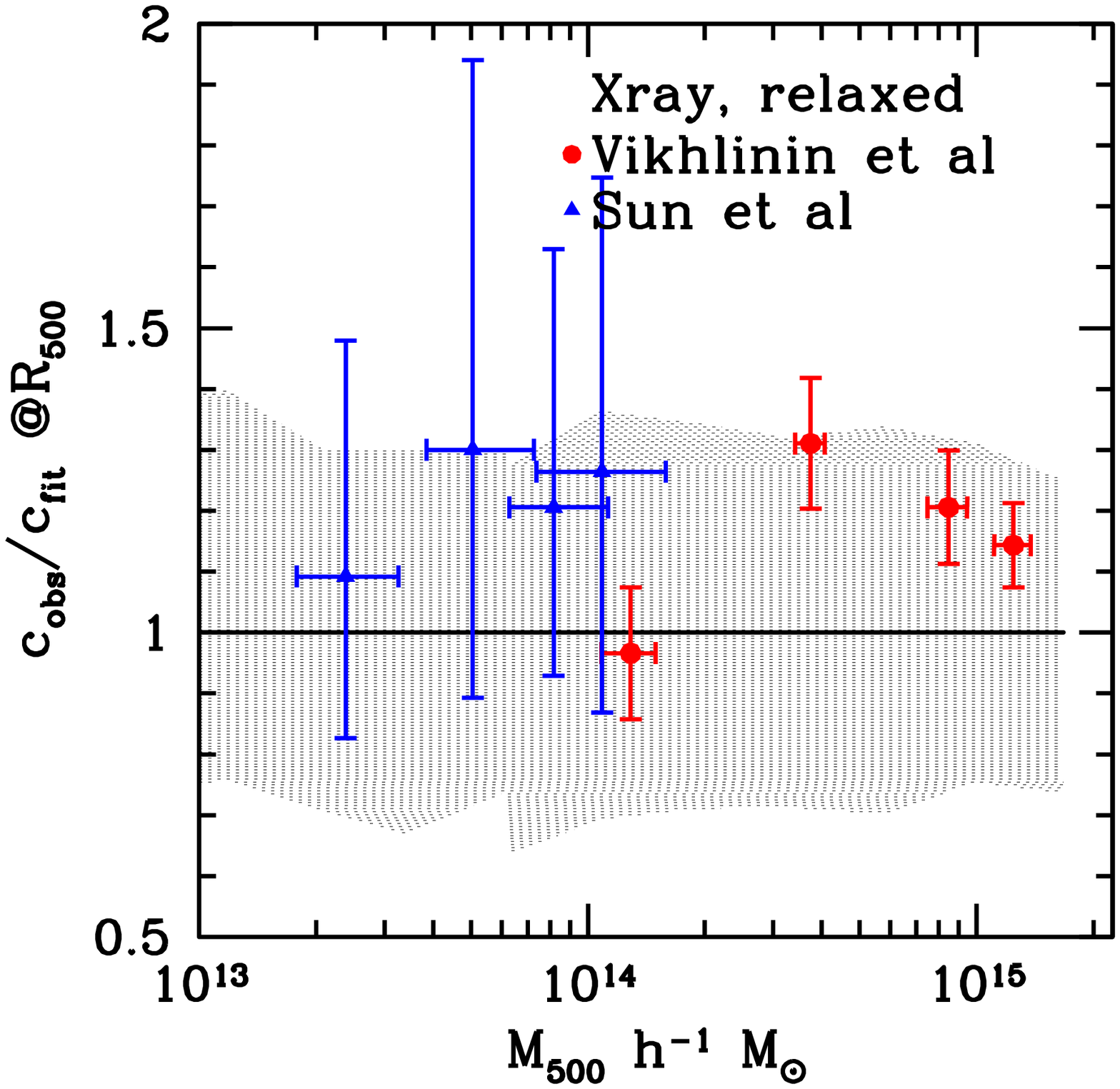}} 
        \end{tabular}
        \caption{Ratio of observed concentration to theoretical
          predictions for relaxed cluster observations. The first two
          sets of data are taken from \cite{schmidt06} and
          \cite{buote06} (left panel, for details see text). The
          shaded area represents the $1\sigma$ boundary for the
          theoretical predictions. The right panel shows the
          comparison against observations of the {\em Chandra} Cluster
          Cosmology Project \citep{vikhlinin09}. This dataset includes
          updates to the results of \cite{vikhlinin05} and adds 6 new
          clusters (Table~\ref{table_vikhlinin}). Note that each data
          point actually represents observations of multiple clusters.
        }
\label{fig:xrayrel}
  \end{center}
\end{figure*}

\begin{figure}
  \begin{center}
    \begin{tabular}{c}
        \resizebox{3.6in}{3.6in}{\includegraphics{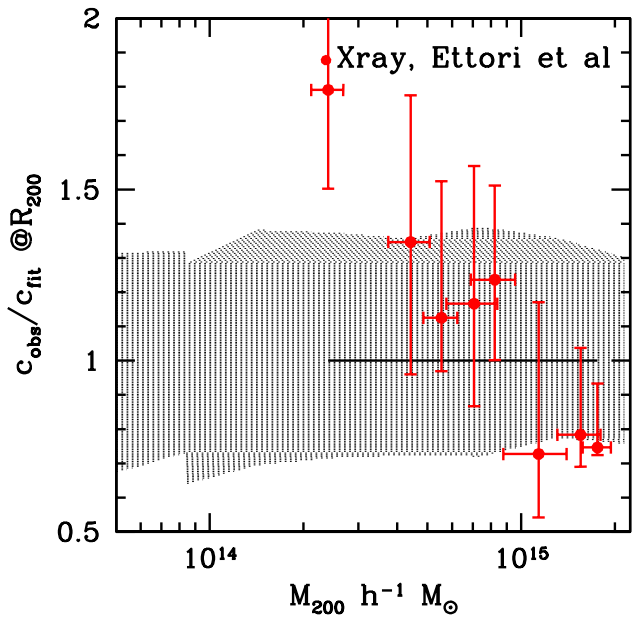}} 
        \end{tabular}
        \caption{Ratio of observed concentration to theoretical
          predictions for the {\em XMM-Newton} cluster observations of
          \cite{ettori11}. The shaded area represents the $1\sigma$
          boundary for the theoretical predictions. Each data point
          actually represents observations of multiple clusters.  }
\label{fig:xray}
  \end{center}
\end{figure}

In this section, we compare our simulation results with some of the
recent observations of the concentration-mass relation for
clusters. The observational results span a variety of techniques,
including strong and weak lensing (e.g., \citealt{comerford07};
\citealt{broadhurst08}; \citealt{mandelbaum08}; \citealt{okabe09};
\citealt{oguri11}) X-ray observations of relaxed clusters (e.g.,
\citealt{buote06}; \citealt{vikhlinin05}; \citealt{schmidt06};
\citealt{vikhlinin09}) and relaxed and unrelaxed clusters
\citep{ettori11}, and cluster kinematics (e.g., \citealt{rines06};
\citealt{wojtak10}). Our aim is to provide a set of figures that
enables the reader to judge by eye the current status of how well the
theoretical predictions match against observations. Because there are
significant observational systematics that are unclear and the
observational statistics are still limited, we do not believe that a
more complete statistical analysis is necessary, or even particularly
useful. The strategy we follow is to take the ratio of each measured
concentration to the theoretically predicted concentration at the
object's observed mass and redshift. We then bin in mass to show a
relatively limited number of comparison points in each figure. Thus
each point in the observation plots represents an average over $\sim$5
data points. (The corresponding Poisson error bars use the improved
formula $\sigma_{\pm}=\sqrt{N_h+1/4}\pm 1/2$ as given by
\citealt{heinrich03}, which asymptotes to $\sqrt{N_h}$ at large
$N_h$.)

We begin our comparison using results from X-ray observations of
relaxed clusters as shown in Fig.~\ref{fig:xrayrel}. \cite{schmidt06}
have measured the concentration of 34 dynamically relaxed clusters
($0.06<z<0.69$) from {\em Chandra} observations (left panel). The
theoretical predictions are in good agreement for masses $M_{vir} >
4\times 10^{14} \mau$, with minor tension at lower masses. The data
presented by \cite{buote06} are a compilation of analyses of relaxed
systems from {\em Chandra} and {\em XMM-Newton}; we show only the
higher mass range, represented by results taken from \cite{point05},
\cite{vikhlinin05}, \cite{zappacosta06}, and \cite{gastaldello07},
spanning a redshift range of $0.016<z<0.23$. All of these results are
in very good agreement with the predictions, lying within the
$1\sigma$ boundary. The right panel shows results from 19 clusters
that were part of the {\em Chandra} Cluster Cosmology Project
\citep{vikhlinin09} ($0.016<z<0.25$), the dataset represented in
Table~\ref{table_vikhlinin}. Once again, the agreement is
excellent. Overall, we conclude that comparisons with X-ray
measurements of relaxed clusters are in good accord with (concordance)
$\Lambda$CDM predictions.

Next we turn to the results of \cite{ettori11} who measured the
concentrations of 44 X-ray luminous clusters ($0.09<z<0.31$) using
{\em XMM-Newton} (Fig.~\ref{fig:xray}). Their sample contains both
relaxed and unrelaxed clusters. As with the \cite{schmidt06}
comparison, we find that the simulation results are in good agreement
with these observations for $M_{200} > 4\times 10^{14} \mau$. As the
authors themselves note, a slope cannot be fitted to their data
because of the narrow mass range of the observations relative to their
errors. Thus, we regard the current level of agreement as being quite
satisfactory. 
\begin{figure*}
  \begin{center}
    \begin{tabular}{cc}
      \resizebox{3.6in}{3.6in}{\includegraphics{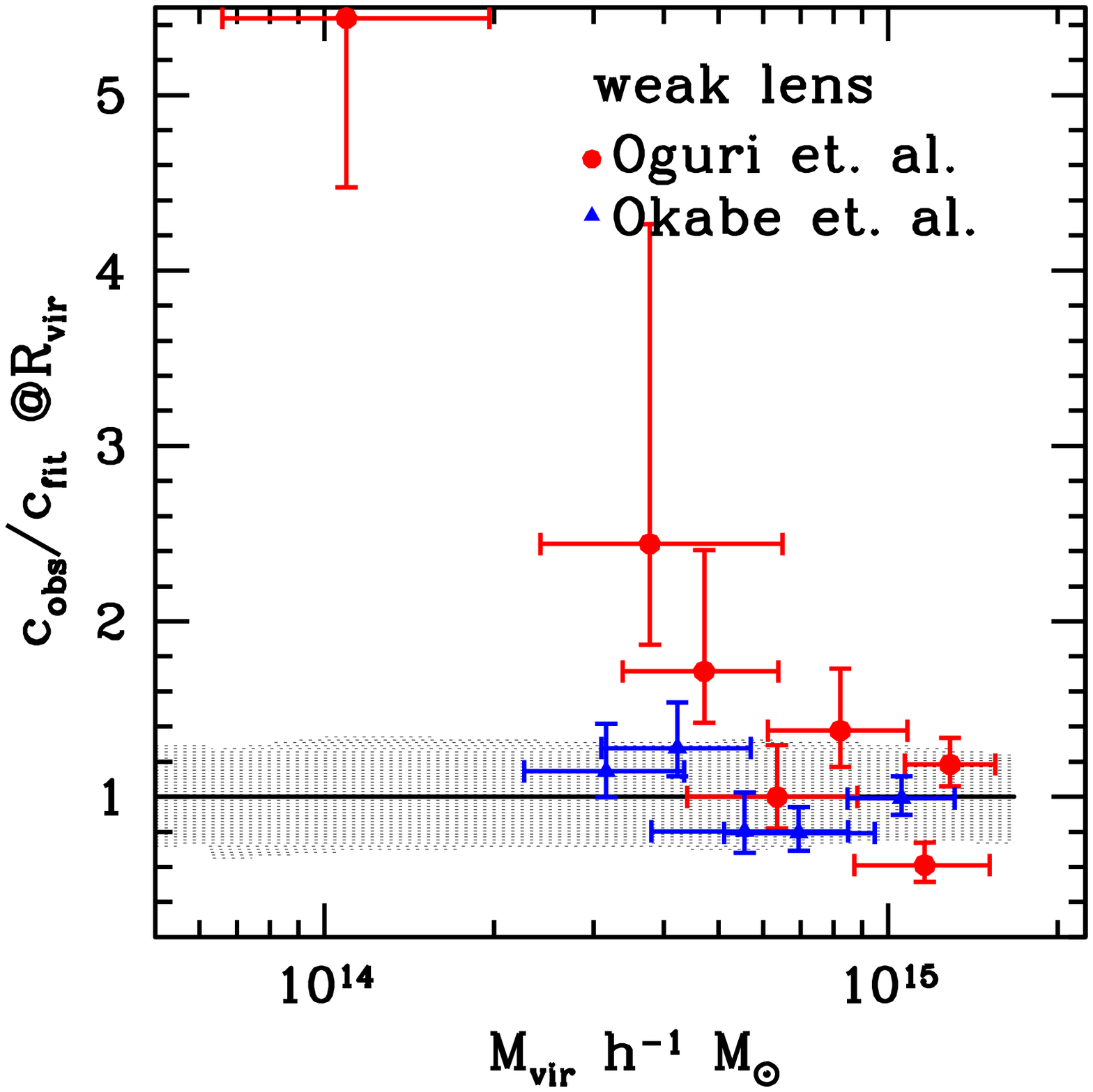}}
         \resizebox{3.6in}{3.6in}{\includegraphics{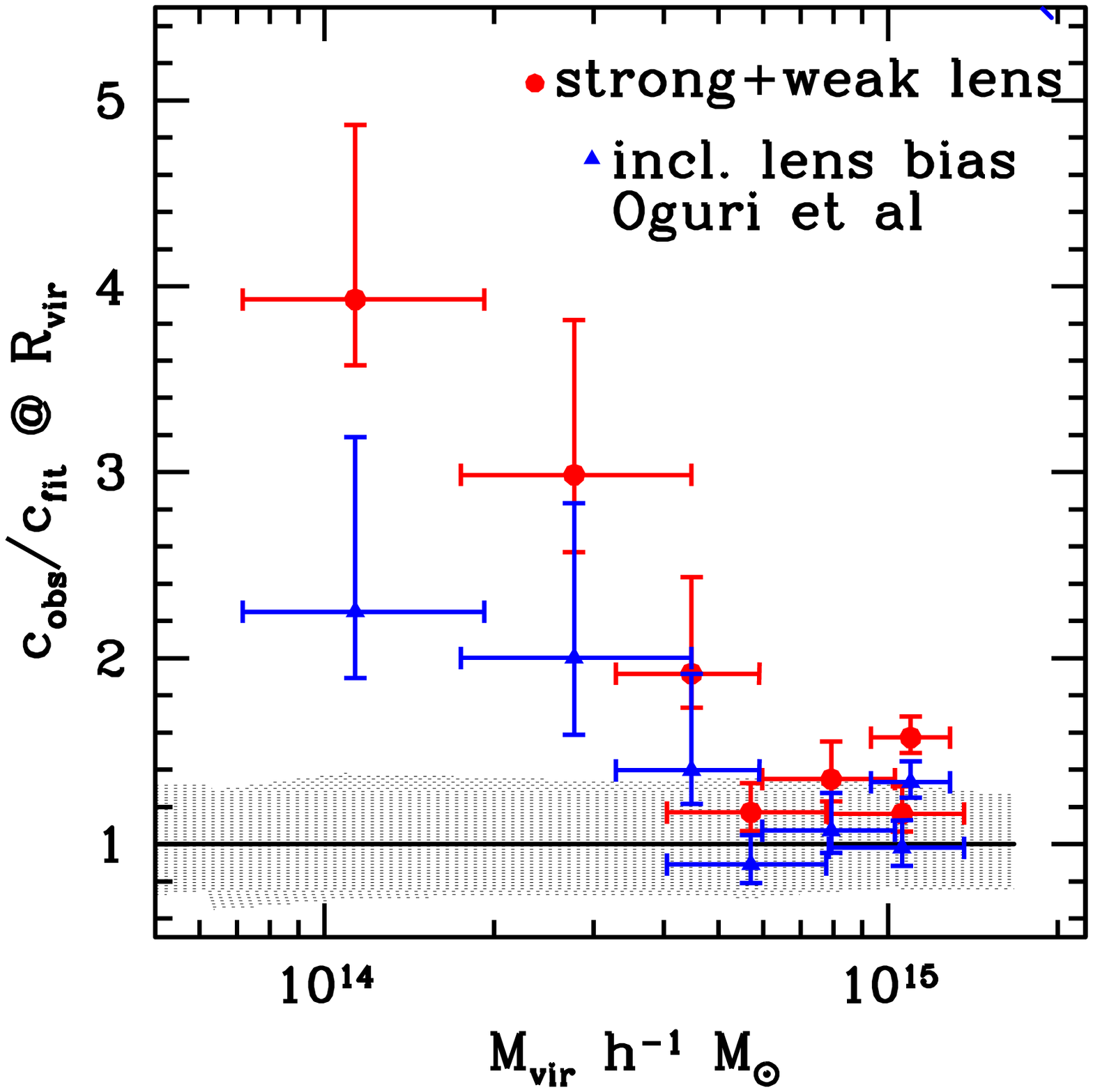}}\\
        \end{tabular}
    \caption{Theoretical versus observed $c-M$ relation for weak and
      strong lensing. The left panel shows weak lensing data from
      \cite{okabe09} and \cite{oguri11}. The \cite{okabe09} results
      are in very good agreements with the predictions while the
      \cite{oguri11} results are strongly discrepant at the low mass
      end. The right panel shows the combined strong and weak lensing
      results from \cite{oguri11} including their bias model-corrected
      prediction (blue). The correction reduces the discrepancy
      significantly but some tension remains below $\sim4\times
      10^{14}\mau$. 
    }
\label{fig:lens}
  \end{center}
\end{figure*}

We now consider lensing measurements of cluster profiles using weak
and strong lensing and combinations thereof. Figure~\ref{fig:lens}
shows the comparison of the theoretical predictions against the
results of LocUss, a weak lensing study of 30 clusters with
Subaru/Suprime-Cam imaging data \citep{okabe09} and a combined strong
and weak lensing analysis of 28 clusters from the Sloan Giant Arcs
Survey \citep{oguri11}. The left panel of Fig.~\ref{fig:lens} shows
the weak lensing results displayed in the same manner as for the X-ray
datasets. The results from \cite{okabe09} are in excellent agreement
with our predictions, completely consistent with the corresponding
measurements from relaxed clusters. The results of \cite{oguri11} are
consistent with our predictions for $M_{vir} > 4\times 10^{14} \mau$,
but at lower masses, there appears to be a significant discrepancy,
with a much steeper $c-M$ dependence. Although baryon cooling may play
a role at smaller masses, there is no convincing reason for such a
large effect -- for which there is no signal in the X-ray data (nor in
the simulations of \citealt{duffy10}). Note that the target selection in
the two surveys is quite different, that of \cite{okabe09} being
essentially volume-limited, while any strong-lensing selected sample
such as that of \cite{oguri11} must have a significant amount of
selection and projection bias (\citealt{rozo08};
\citealt{meneghetti10}). Note also that an analysis based on mock weak
lensing observations in the MS \citep{bahe11} has shown that there is
bias in weak lensing measurements of concentration as well, tending to
depress the measured concentration by a small amount from the
predicted value.

The right panel of Fig.~\ref{fig:lens} shows the combined strong plus
weak lensing analysis including a model for lensing bias
\citep{oguri11}. Processing the results through the lensing bias model
(by enhancing the theoretical prediction) brings down the discrepancy
significantly but there is still evident tension for masses $M_{vir} <
4\times 10^{14} \mau$. Nevertheless, we note that there is a clear
trend of lensing concentrations reducing over time and becoming more
consistent with the theoretical predictions. Other data our results
appear to be in agreement with can be found in \cite{comerford07}
(strong lensing) and \cite{coe12} (strong and weak lensing). A
cautionary note regarding weak lensing concentration measurements of
clusters is provided in Figure~6 of \cite{comerford07} regarding
Abell~1689 and in the results given in \cite{israel11} as part of the
{\em 400d} weak lensing survey.

Instead of using individual objects, a stacked statistical analysis
can be applied to clusters, as carried out using the Sloan Digital Sky
Survey by \cite{mandelbaum08}, to a cluster mass range of $\sim
6\times 10^{14} \mau$. This analysis sees no evidence for a major
boost in concentration at lower masses and the final result --
$c_{200b} \sim 4.6 \pm 0.7$ at $\left < z \right> =0.22$ at a mass of
$M_{200b} \sim 10^{14} \mau$ is $20-40\%$ less than our prediction of
$c_{200b}\sim 6.5$ at the corresponding mass. The mild $c-M$
dependence they observe is however in good agreement with our
predictions -- $\sim 0.09$ compared to the observed slope of $0.13 \pm
0.07$.

\begin{figure}
  \begin{center}
    \begin{tabular}{c}
      \resizebox{3.6in}{3.6in}{\includegraphics{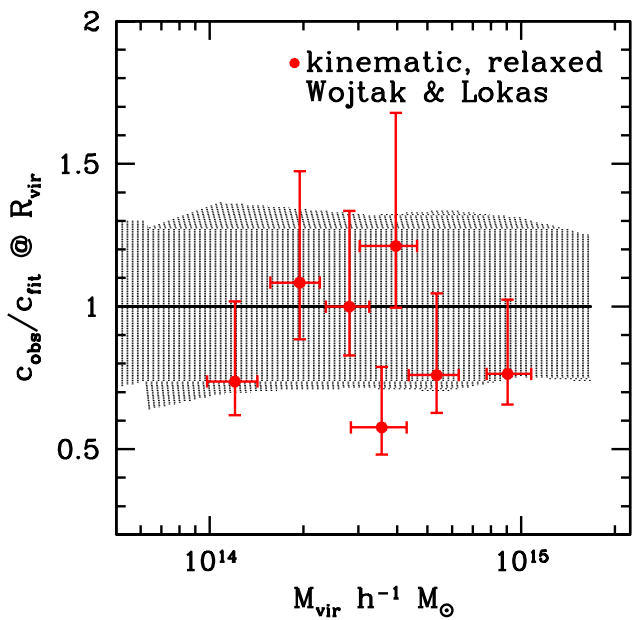}}
        \end{tabular}
    \caption{Ratio of measured to predicted concentrations; the data
      is taken from the results of a projected phase space analysis by
      \cite{wojtak10}. 
    }
\label{fig:kin}
  \end{center}
\end{figure}

Finally, we consider the estimates of the concentration using galaxy
kinematics in clusters. \cite{rines06} matched X-ray cluster catalogs
with SDSS and used infall patterns to compute cluster mass
profiles. The $c_{200}$ concentration has significant scatter --
values ranging from $5-17$ -- but their best-fit average profile, with
fits restricted to $r\leq R_{200}$, yields $c_{200}=5.2$ which, at an
average mass of $M_{200}\simeq 10^{14} \mau$, is in agreement with our
predictions. More recently \cite{wojtak10} have published an analysis
of 41 relaxed galaxy clusters ($0.013<z<0.095$); we compare our
predictions with their results in Fig.~\ref{fig:kin}. Although their
data has considerable scatter it is in quite reasonable agreement with
the predictions from simulations. Thus, despite possible systematic
difficulties with such methods (see, e.g., \citealt{white10}), the
current results are in reasonable accord with theoretical
expectations.

\section{Summary and Discussion}
\label{section:disc}

We presented results for the concentrations of dark matter halos using
a set of large-volume simulations. With a total volume roughly 1-2
orders of magnitude larger compared to previous simulations, we
focused on studying the $c-M$ relation for massive clusters. As shown
in the past, at the high mass end, the $c-M$ relation becomes flatter
at $z=0$ and the flattening becomes more significant at higher
$z$. The mean concentration of the sample when expressed in terms of
the peak height parameter, $\nu(M,z)=\delta_c/\sigma(M,z)$, shows a
roughly uniform slope at all redshifts. Indeed, the slope of the
$c-\nu$ relation does not change with redshift. The amplitude of the
$c-\nu$ relation evolves by about $30\%$ at the high mass end from
$z=0-2$. The $z$-evolution is consistent with the results of
\cite{gao07}, although the overall amplitude of the concentration
differs because of the different choices of $\sigma_8$. We do not
observe a rise in concentration at higher masses as reported by
\cite{klypin10} and \cite{prada11} (the Appendix includes further
discussion). Because of our large halo sample, we can study the
distribution of the concentration in individual mass bins; we find
that the distribution of concentrations is well-described by a
Gaussian PDF \citep{lukic09,reed11}. Thus the halo profile shape can
be described by two parameters -- the mean concentration and its
standard deviation. By comparing results across a number of $w$CDM
cosmologies, we find that the standard deviation is roughly universal,
$\sigma_c=0.33 c$, and does not change with redshift, halo dynamical
state, or cosmological parameters.

We investigated how the concentration changes as the cosmological
parameters are varied using a set of 18 runs spanning the $w$CDM
cosmology parameter space. The parameter range covers the 2$\sigma$
variation around the best fit WMAP5 cosmology. We find that over this
parameter range, the $c-M$ relation varies by $\sim \pm$ 20\%,
although the standard deviation $\sigma_c$ follows the relation
$\sigma_c=0.33 c$. As suggested by our work on the $w$CDM models, and
also previous studies of redshift evolution, the halo formation epoch,
and hence the concentration, depends on the matter fluctuations, slope
of the power spectrum and the growth factor. Thus calibrating the
$c-M$ relation as a function of cosmology is important for a wide
variety of problems, ranging from galaxy formation, the weak lensing
shear power spectrum, to the case of assembly bias in clusters. We
will address the cosmology dependence of the $c-M$ relation in detail
using more simulations and analytical models elsewhere.

The simulation predictions are in good agreement with observations
from strong lensing, weak lensing, galaxy kinematics, and X-ray data
for massive clusters with masses $M_{vir} > 4\times 10^{14} \mau$. At
lower masses, different observations suffer from different sources of
systematic error. For example, the lensing data need to account for
bias due to the triaxiality of halos while the X-ray data typically
ignore the non-thermal pressure component in galaxy clusters which can
lead to a systematic underestimate of the cluster mass
\citep{lau09}. The simulations, on the other hand also need to account
for baryonic effects which play a bigger role as the halo mass
decreases. As a result, due to cooling, gravity-only simulations may
predict $20-30\%$ lower concentration for clusters with masses
$M_{vir} < 4\times 10^{14} \mau$. The fact that most recent
observations are in agreement with the simulation results (and amongst
themselves) to better than $20\%$ for massive clusters ($M_{vir} >
4\times 10^{14} \mau$) indicates that baryonic effects influencing the
cluster mass profile are indeed small and that the individual
observational systematics are under some level of control.

\appendix

\section{Systematics and Robustness checks}
\label{sec:checks}

\begin{figure*}
  \begin{center}
    \begin{tabular}{cc}
      \resizebox{3.6in}{3.6in}{\includegraphics{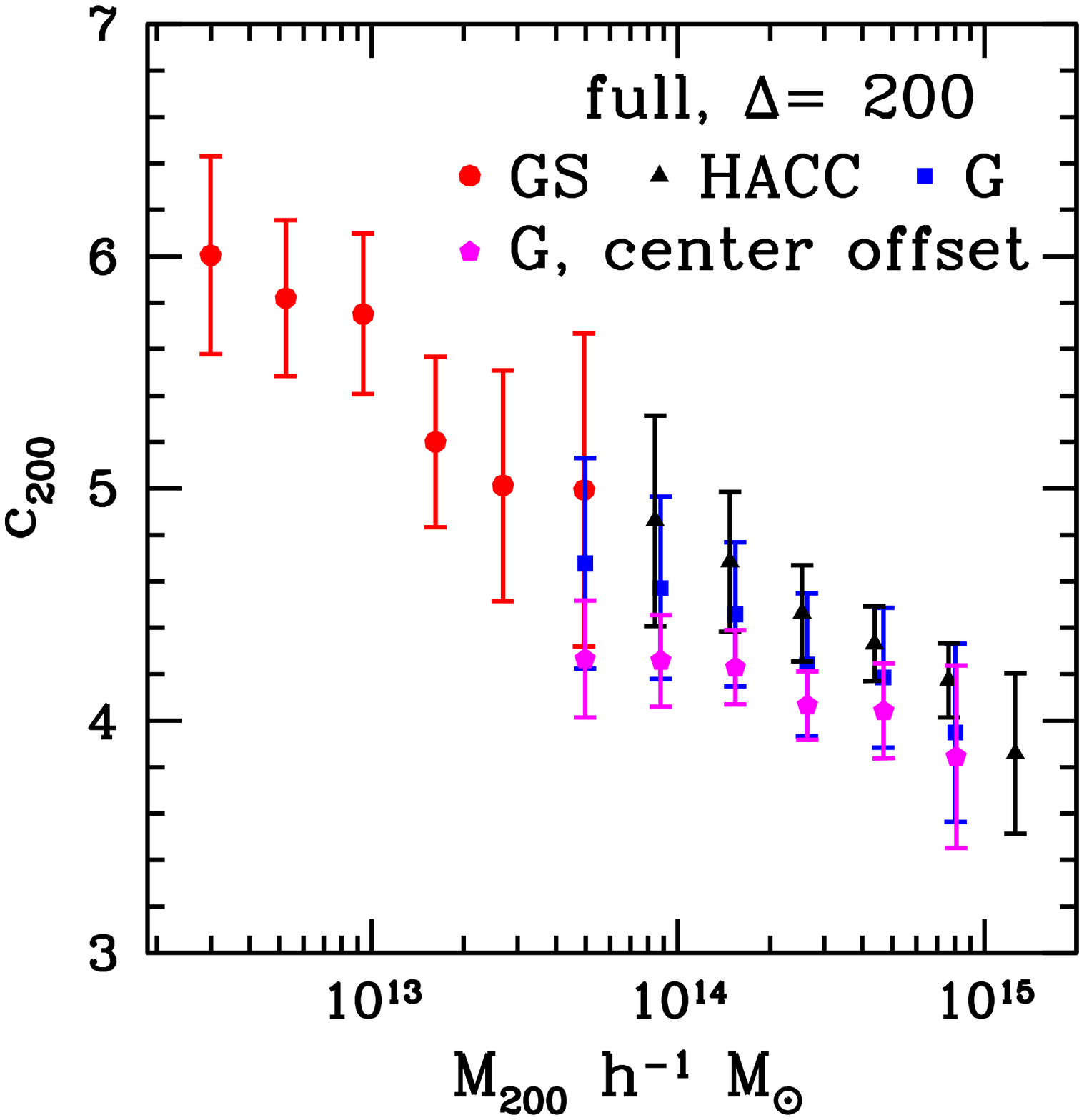}}
      \resizebox{3.6in}{3.6in}{\includegraphics{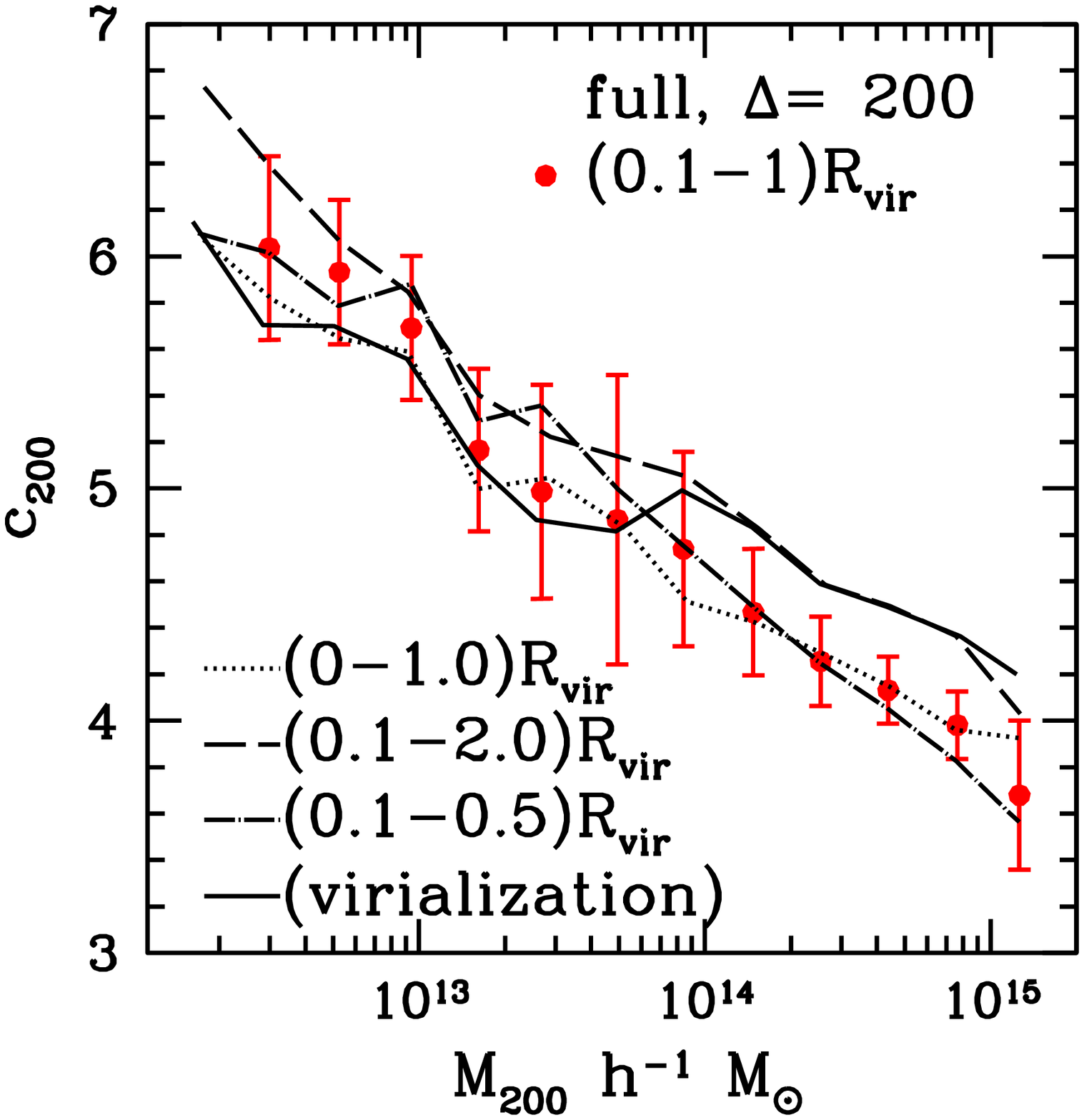}} 
     \end{tabular}
     \caption{Tests to identify possible sources of systematic
       error. The left panel shows the three runs used in this paper
       and how the force resolution affects concentration
       measurements. There is excellent agreement between the GS and
       HACC runs, even though they were run using two completely
       different codes with different settings. The lower-resolution G
       run (resolution$=35h^{-1}$kpc) is systematically lower by
       about $5\%$ compared to the HACC run
       (resolution$=7h^{-1}$kpc). We also study the possible effect of
       a misestimate of the halo center for the G run which can lead
       to a further reduction of concentration values. The right panel
       studies two other systematics issues, (i) the profile range
       used for fitting, and (ii) effect on concentration measurements
       when halos with large radial infall velocity are removed from
       the sample (solid curve). See the text for further discussion.}
\label{fig:cmsys}
  \end{center}
\end{figure*}

\begin{figure*}
  \begin{center}
    \begin{tabular}{ccc}
      \resizebox{2.5in}{2.5in}{\includegraphics{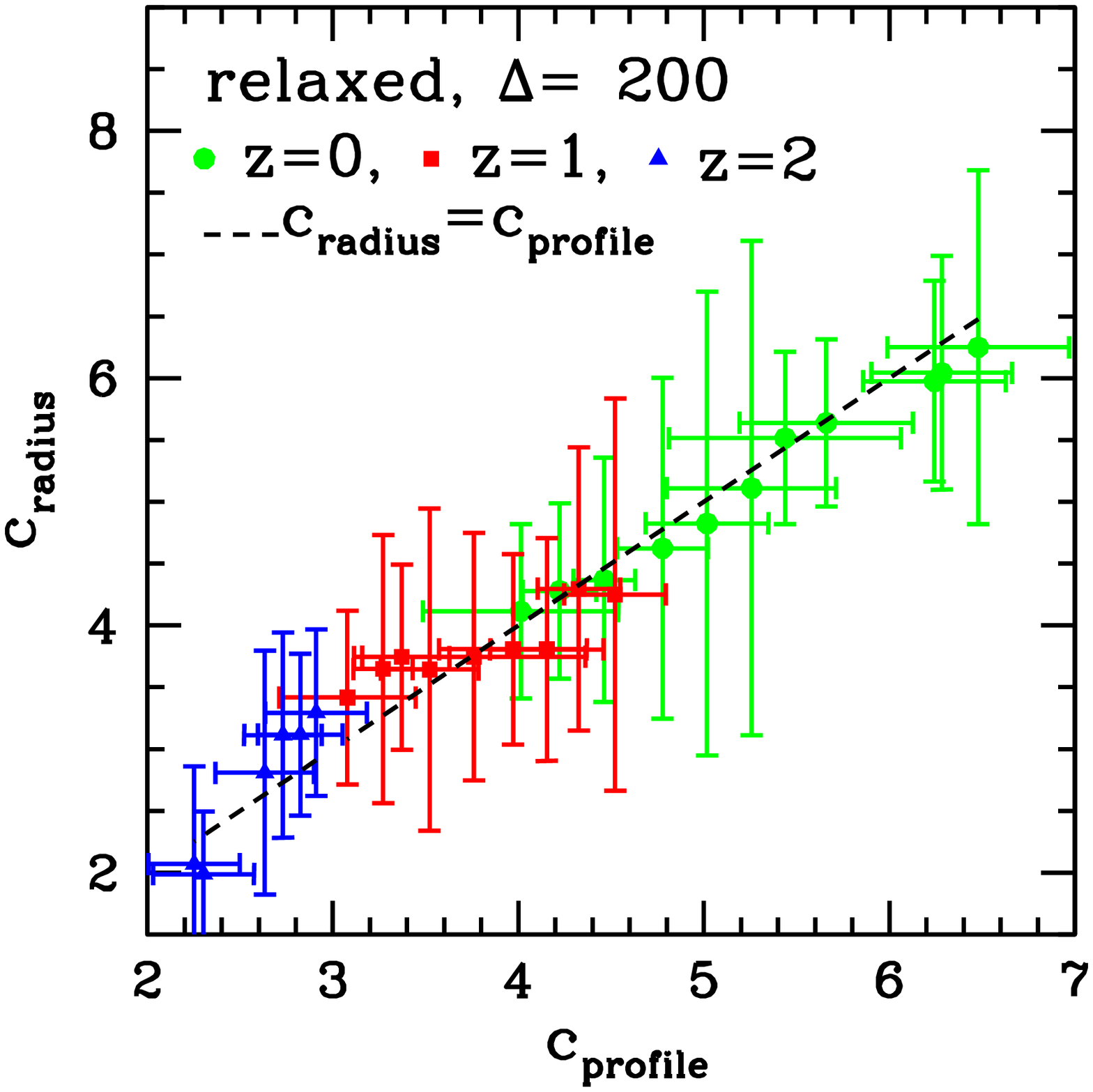}}  
  \resizebox{2.5in}{2.5in}{\includegraphics{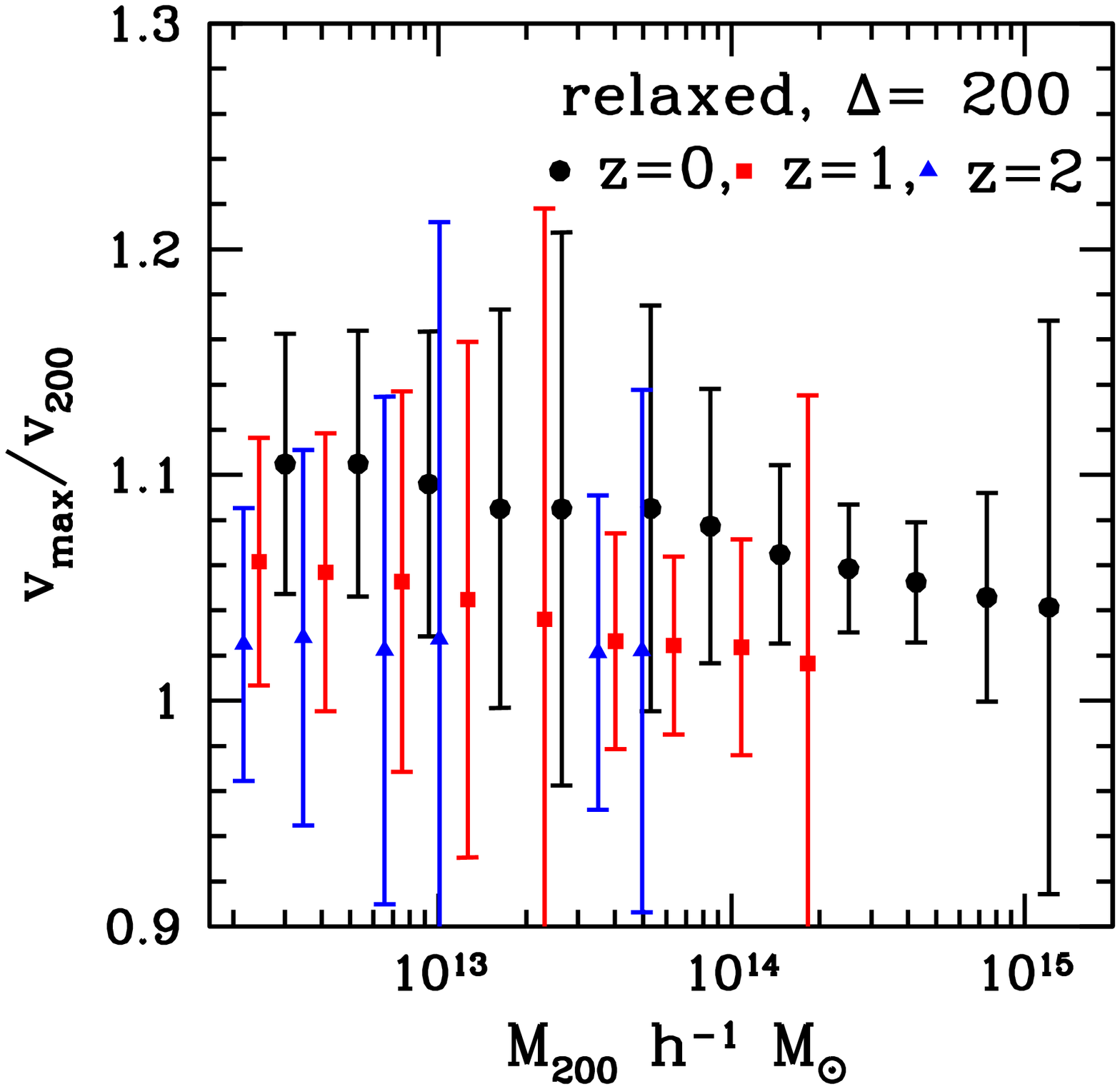}}
   \resizebox{2.5in}{2.5in}{\includegraphics{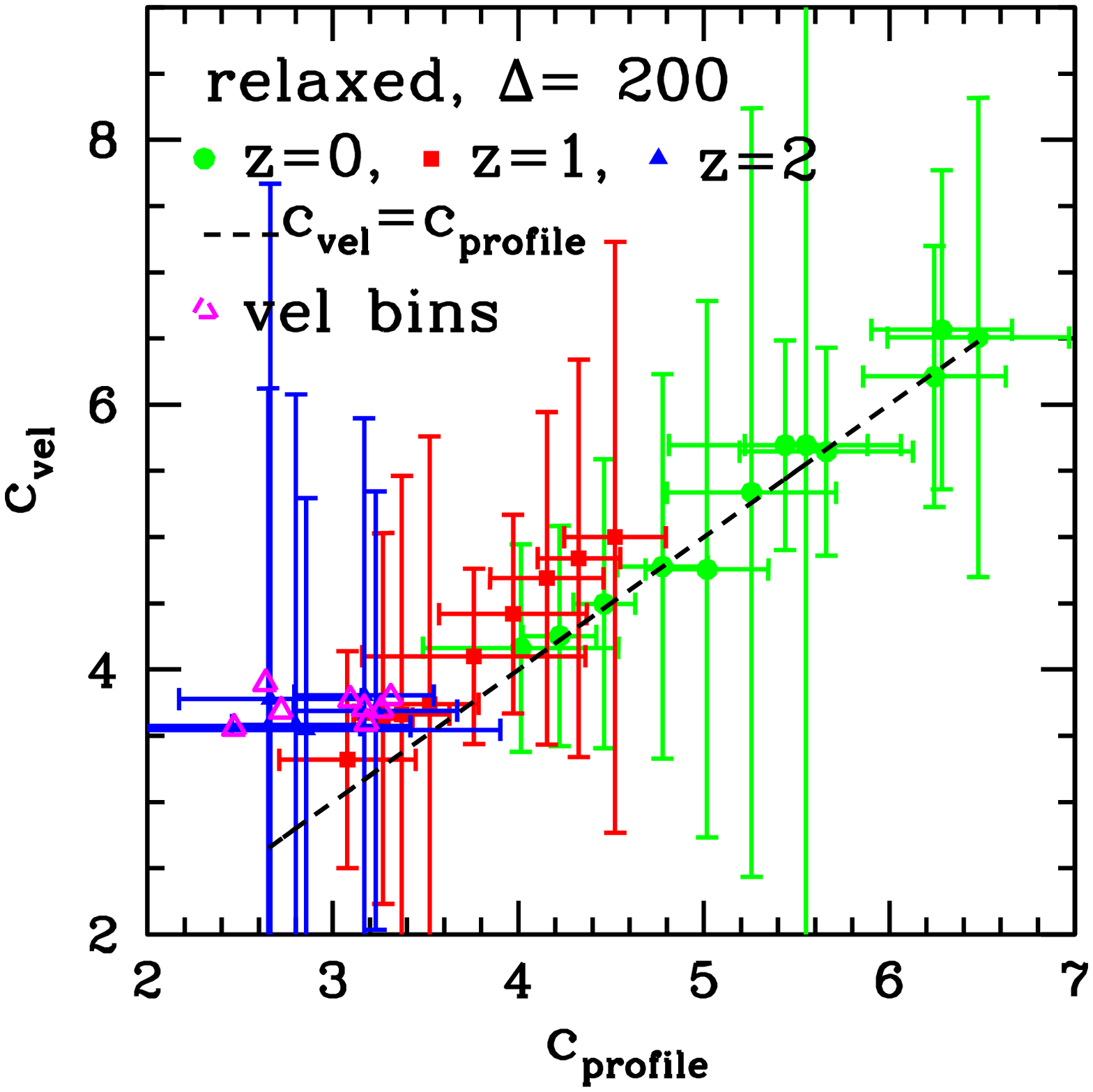}}
     \end{tabular}
     \caption{Comparison (left and right panel) between $c_{200}$
       measured for relaxed clusters using profile fitting and that
       obtained from the radius ratio (left panel) and
       $v_{max}/v_{200}$ (right panel). The diagonal line represents
       the ideal case when the measured concentrations agree exactly.
       The middle panel shows the ratio of the maximum circular
       velocity to that at radius $R_{200}$ for $z=0$ for the relaxed
       sample as a function of mass. Note the smooth cross-over
       between the GS and HACC runs at
       $M_{200}\sim10^{14}h^{-1}$M$_{\odot}$ at redshifts $z=0,1$
       (Cf. Fig.~\ref{fig:c-M}).  }
\label{fig:cfit-cother}
  \end{center}
\end{figure*}

In this Appendix, we investigate various sources of possible
systematic errors in determining halo concentrations. We note that the
simulations were carried out keeping in mind error control
requirements on force-resolution, time-stepping errors, and mass
resolution that have been spelled out in the literature
\citep{tormen96, power03, heitmann08, lukic07}.

We locate halo centers by using a very fast FOF method. In principle,
there is a very mild selection effect induced by the choice of linking
length -- if the FOF finder links two halos, then only the higher
density center of one halo will be used, and the other halo will be
statistically lost (see, e.g. Fig.~12 in \citealt{lukic09}). A smaller
linking length would in effect free up the other halo as well,
although it would slow down the center-finding algorithm. In practice,
however, this is not an issue given the fact that individual
concentrations need to be extracted with small errors. To do so we use
a minimum number of $\sim 2000$ particles per halo; tests have shown
that with more than $500$ particles per halo, there is excellent
agreement between our method of halo finding and conventional
spherical overdensity halo finders~\citep{knebe11}.

Another problem with halo center finding is miscentering, which in
general will tend to reduce the concentration. To produce a
quantitative estimate for this effect, we consider an extreme example
by offsetting the center of every halo by $30h^{-1}$kpc (a value of
the same order as the force resolution) and recomputing the profile of
every halo in the G run. As shown in Figure~\ref{fig:cmsys}, left
panel, randomly offseting the center by this amount reduces the $c-M$
relation by about 10\% at $M_{200}=5\times 10^{13} h^{-1}M_\odot$ with
the difference vanishing towards the high mass end. (This sort of
misestimation is more relevant to analyses with stacked halos.)

For profile tests across the different runs, we begin by comparing the
GS and HACC runs (left panel of Figure~\ref{fig:cmsys}). Note that
although these runs have been carried out with completely different
codes, the data for $c_{200}$ goes over smoothly from one mass range
to another. ({\sc GADGET-2} and HACC were run with roughly similar
force resolution.) The G runs were originally carried out for a
different purpose, hence their force resolution is somewhat lower. As
expected, this has the consequence of mildly reducing the
concentration \citep{tormen96, power03} by about $5\%$.

We have tested our profile-fitting method by generating Monte Carlo
NFW profile samples using different numbers of particles; with the
particle numbers used to sample halos kept larger than 2000, the
method was accurate to a few percent (worst case) and superior to
simpler methods based on radius ratios and variants thereof. To
investigate how other parameters could affect concentration values, we
went back to using the halos from the simulations.  The right panel of
Figure~\ref{fig:cmsys} shows the effect of varying the range of the
halo profile used to fit to the NFW form. Changing the starting radius
from $r=0.1 R_{vir}$ to $r=0$, with the outer limit fixed at
$r=R_{vir}$ reduces the overall $c-M$ relation by about 5\%
(resolution/particle undersampling limitations). Fixing the starting
radius at $0.1 R_{vir}$ but changing the stopping radius to $2
R_{vir}$, only changes the relation by a negligible amount from the
fiducial range of $(0.1-1)R_{vir}$. Reducing the stopping radius to
$0.5 R_{vir}$ steepens the $c-M$ relation by about 10\%.

Because our primary interest is in halos that have mass significantly
in excess of $M_*$, it is important to ask what possible systematic
effects could arise from fitting such objects without paying attention
to their infall structure. The average radial velocity of a halo
fluctuates around zero out to an infall radius, $r_{inf}$, beyond
which it goes negative, this transition roughly defining the boundary
of the infall region. Purely as an informal nomenclature, we refer to
the region internal to the infall radius as the virialized region. We
find all halos that have $R_{vir}>r_{inf}$ and exclude them from the
analysis, thus focusing attention on halos that have much less infall
contamination. The result is shown by the solid line in the right
panel of Figure~\ref{fig:cmsys}. Not unexpectedly, cluster size halos
with masses $\sim10^{14}h^{-1}$M$_{\odot}$ and greater are much more
sensitive to this cut, and display an enhancement of the $c-M$
relation by about $10\%$ when only the `virialized' sub-sample is
used.

\begin{figure*}
  \begin{center}
    \begin{tabular}{cc}
      \resizebox{2.5in}{2.5in}{\includegraphics{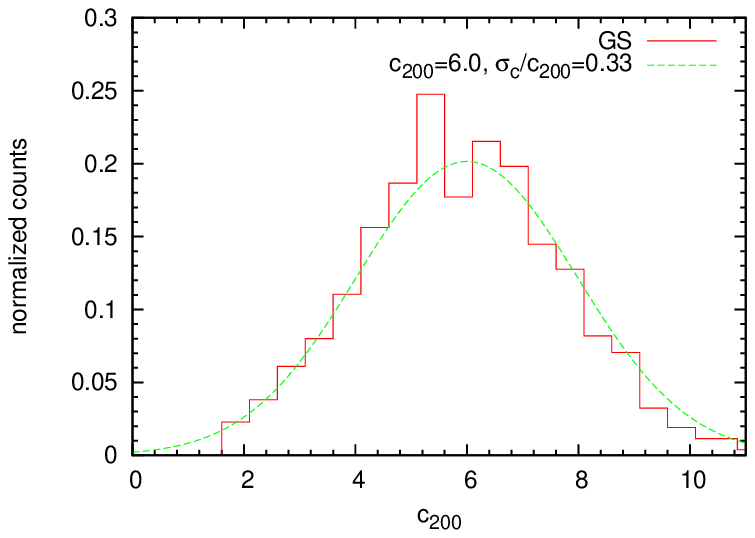}}
         \resizebox{2.5in}{2.5in}{\includegraphics{c-M.hist.HACC.eps}}
        \resizebox{2.5in}{2.5in}{\includegraphics{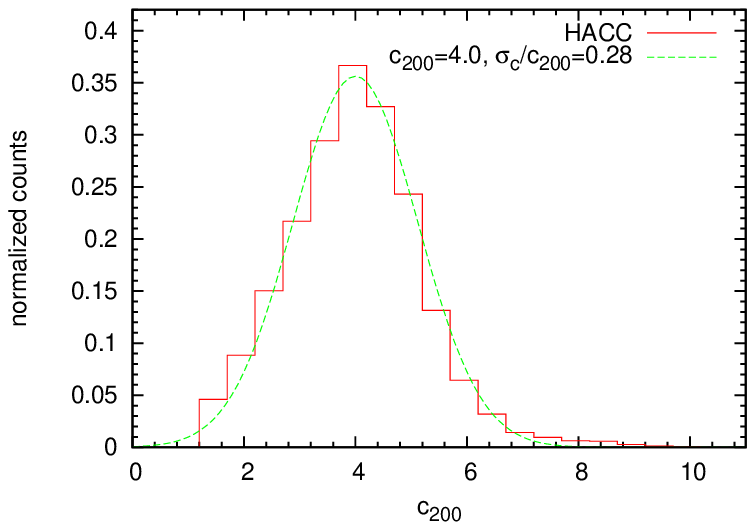}} 
       \end{tabular}
       \caption{$c-M$ distribution at three mass
         bins-$M_{200}=5\times10^{12} h^{-1} M_\odot$ (left from the
         GS run), $1.5\times 10^{14} h^{-1} M_\odot$ (middle, HACC)
         and $8\times 10^{14} h^{-1} M_\odot$ (right, HACC) from the
         halos drawn from the full sample at $z=0$. Lines show the
         Gaussian distribution with standard deviation $\sigma_c/c
         \sim 0.33$.}
\label{fig:cMdist}
  \end{center}
\end{figure*}

As mentioned in Section~\ref{section:halo}, the halo concentration can
be measured in different ways, even if one sticks to the NFW
definition(s) of concentration. Therefore, it is important to
investigate what sources of uncertainty can arise from using different
definitions that may be mathematically equivalent, but not
operationally the same. Here we study two alternative independent
techniques for measuring concentrations -- the radius ratio and the
maximum circular velocity. The radius ratio method is very simple: We
measure the radius at $\Delta=300$ and $200$ for each halo in our
sample. Then, assuming the halos are described by an NFW profile, the
radius ratio is given by
\begin{equation}
\frac{300R^3_{300}}{200R^3_{200}}= \frac{m(c_{200}R_{300}/R_{200})}{m(c_{200})}
\label{eq:rratio}
\end{equation}
where $m(x)= \ln(1+x)-x/(1+x)$. Given a measurement of
$R_{300}/R_{200}$, one solves the nonlinear equation
Eq.~\ref{eq:rratio} for $c_{200}$. The mean $c-M$ relation is then
obtained in the same way as for profile fitting. The left panel of
Figure~\ref{fig:cfit-cother} shows the concentration measured using
the profile fit and by the radius ratios. For this test we focused on
the relaxed sample, although the full sample gave identical
results. The diagonal line in the figure denotes the ideal exact
agreement of the concentration measures from the profile fit and from
the radius ratio. As the results in the figure show, the two methods
agree quite well within the specified errors. The error in the
concentration measurement using the radius ratio arises from the error
in determining the radii -- the Poisson error from the total number of
particles inside $R_{300}$ and $R_{200}$. As expected, the radii are
determined quite accurately as there are large number of particles
inside these radii, but because of the logrithmic nature of
Eq.~\ref{eq:rratio}, the $\Delta c_{200}$ become non-negligible. The
error on the mean concentration also includes the Poisson error due to
the finite number of halos in the individual mass bins.

The second method we investigate relies on using a proxy for the
maximum circular velocity of a halo~\citep{klypin10, prada11}. The
circular velocity is given by
\begin{equation}
v^2=GM(< r)/r
\label{eq:vcirc}
\end{equation}
For each halo in the sample we determine the maximum value of
$v^2_{max}=max[GM/r]$ indirectly, by using the halo's mass profile to
determine the RHS of Eq.~\ref{eq:vcirc}, and then divide by
$v^2_{200}= GM_{200}/R_{200}$. Assuming an NFW form, one can relate
$v^2_{max}/v^2_{200}$ to the concentration, $c_{200}$,
\begin{equation}
\frac{v^2_{max}}{v^2_{200}}=\frac{0.2162 c_{200}}{m(c_{200})}
\label{eq:vratio}
\end{equation}
and solve for $c_{200}$. The middle panel of
Figure~\ref{fig:cfit-cother} shows the ratio $v_{max}/v_{200}$ as a
function of $M_{200}$ for three redshifts for the relaxed cluster
sample. The right panel compares the concentrations obtained from the
maximum velocity method and profile fitting. Again, the methods agree
quite well within the error estimates, the velocity method being
noisier. The middle panel of Figure~\ref{fig:cfit-cother} shows the
ratio $v_{max}/v_{200}$ as a function of mass. Note that we cross over
very smoothly from the GS run to the HACC run at
$M_{200}=10^{14}h^{-1}$M$_{\odot}$, more evidence for an excellent match
between the results from these two simulations. One problem with this
method is that because $r_{max}\sim 2.2 r_s$ (where $r_{max}$ is the
radius where $v$ reaches $v_{max}$), at low concentrations, $r_{max}$
becomes very close to $r_{\Delta}$, and is therefore very sensitive to
any noise in the data, which will (i) result in biasing the
concentration to a higher value (as seen in the flattening of the data
at $c\sim 3$ in Fig.~\ref{fig:cfit-cother}, right panel), and (ii)
make the result very sensitive to the shape of the measured profile at
$r\sim r_{\Delta}$, increasing the possibility of systematic
errors. Finally, both \cite{klypin10} and \cite{prada11} have found
that the maximum velocity method (along with their halo selection)
leads to an upturn in the $c-M$ relation at the high mass end at
higher redshifts. We are unable to confirm this effect in our
measurements, where we can investigate it (at redshifts, $z=1,2$).

We provide more information regarding the distribution of the
concentrations, $c_{200}$, for the full halo sample at $z=0$ by
considering three representative mass bins, $M_{200}=5\times10^{12}
h^{-1} M_\odot$, $1.5\times 10^{14} h^{-1} M_\odot$ and $8\times
10^{14} h^{-1} M_\odot$, as shown in
Figure~\ref{fig:cdist_gauss}. Previous studies \citep{jing00, shaw06,
  neto07, duffy08}, have fitted the concentration distribution to a
log-normal distribution. However, this distribution is also very well
described by a Gaussian as noted by \cite{lukic09} and
\cite{reed11}. Figure~\ref{fig:cdist_gauss} shows that a Gaussian
distribution provides a very good fit to our data, with relatively
negligible non-Gaussian tails.

We conclude that while statistical errors on the concentration-mass
relation may have achieved $\sim 5\%$ accuracy in recent simulation
studies, systematic uncertainties of the order of $10\%$ are
apparently difficult to avoid.

\acknowledgments

It is a pleasure to acknowledge discussions and collaborations with
Joanne Cohn, Zarija Luki\'c, Darren Reed, Alexey Voevodkin, and Martin
White. We acknowledge several motivating conversations with Masahiro
Takada (SH) and Mike Gladders (SH and KH). We thank Volker Springel
for discussions and for making {\sc GADGET-2} publicly available. We
are indebted to Patricia Fasel and Adrian Pope for their contributions
to the HACC analysis framework. We thank Anatoly Klypin and Francisco
Prada for pointing out a numerical error in an earlier version of the
Appendix (now corrected) and for discussions of results obtained by
our two groups. A special acknowledgment is due to the resource
allocation awarded to us on the hybrid supercomputer Cerrillos and
other clusters under the Los Alamos National Laboratory Institutional
Computing initiative. Part of this research was supported by the DOE
under contract W-7405-ENG-36.  The authors acknowledge support from
the LDRD programs at Los Alamos National Laboratory and Argonne
National Laboratory, where analysis was performed on the Eureka
cluster. SB and KH were supported in part by NASA.

\end{document}